# Bio-inspired site characterization - towards soundings with lightweight equipment

*Alejandro* Martinez[1#], *Yuyan* Chen[2], and *Riya* Anilkumar[1]

[1]*University of California Davis, Department of Civil and Environmental Engineering, 2001 Ghausi Hall, Davis, CA, USA*
[2]*Tianjin University, National Facility for Earthquake engineering Simulation, No. 135 Yaguan Road, Tianjin, China*
[#]*Corresponding author: amart@ucdavis.edu*

**ABSTRACT**

Equipment used for site investigation activities like drill rigs are typically large and heavy to provide sufficient reaction mass to overcome the soil's penetration resistance. The need for large and heavy equipment creates challenges for performing site investigations at sites with limited accessibility, such as urban centres, vegetated areas, locations with height restrictions and surficial soft soils, and steep slopes. Also, mobilization of large equipment to the project site is responsible for a significant portion of the carbon footprint of site investigations. Successful development of self-burrowing technology can have enormous implications for geotechnical site investigation, ranging from performance of in-situ tests to installation of instrumentation without the need of heavy equipment. During the last decade there has been an acceleration of research in the field of bio-inspired geotechnics, whose premise is that certain animals and plants have developed efficient strategies to interact with geomaterials in ways that are analogous to those in geotechnical engineering. This paper provides a synthesis of advances in bio-inspired site investigation related to the (i) reduction of penetration resistance by means of modifying the tip shape, expanding a shaft section near the probe tip, applying motions to the tip like rotation and oscillation, and injecting fluids and (ii) generation of reaction forces with temporary anchors that enable self-burrowing. Examples of prototypes that have been tested experimentally are highlighted. However, there are important research gaps associated with testing in a broader range of conditions, interpretation of results, and development of hardware that need to be addressed to develop field-ready equipment that can provide useful data for geotechnical design.

## 1. Introduction

### 1.1. Motivation

Soils such as dense sands and gravels, highly overconsolidated clays, and cemented layers have large penetration resistances, creating challenges for site characterization activities. Soil penetration is an energy-intensive process that is usually accomplished through direct pushing, impact driving, or excavation. Fig. 1 shows equipment typically used in geotechnical engineering activities to advance probes for site characterization, install piles, and excavate tunnels. In all these cases, the equipment is large in size, resulting in mobilization costs and logistics challenges that can have important impacts on engineering projects.

The performance of in-situ tests that require inserting probes in the ground, such as the Cone Penetration Test (CPT), Dilatometer Test (DMT), Pressuremeter Test (PMT), and the installation instrumentation for monitoring sites, such as pore pressure transducers, inclinometers, and accelerometers, rely on penetrating or excavating soils. Thus, they require equipment that has enough reaction mass to overcome the soil's resistance. For example, typical rigs used for CPT and DMT testing have masses ranging from 15 to 40 tons in order to reduce the likelihood of refusal that would result in termination of the sounding before the desired depth is reached. Use of such large equipment requires proper access to the site in the form of permanent or temporary roads or cleared areas, which can be a challenge in dense urban centres, vegetated areas, locations with height restrictions and soft surficial soils, and steep slopes. For example, temporary embankments must be built on the face of dams to provide access to rigs if the underlying soils are to be characterized or the dam instrumented. In addition to the economic impacts, mobilization of large rigs to project sites can be responsible for a significant portion of the carbon footprint. For example, over 50% of the energy spent in typical site investigation activities, such as a 30 m deep CPT sounding requiring mobilization of a 20-ton rig for 160 km, comes from mobilization of the rig (Purdy et al. 2022). Thus, reducing the external vertical reactions necessary to penetrate soils would allow for use of smaller equipment, which could make site investigation activities simpler and more efficient. This can be done either by reducing the penetration resistance of soils or by generating the required reaction force on-site through temporary anchors.

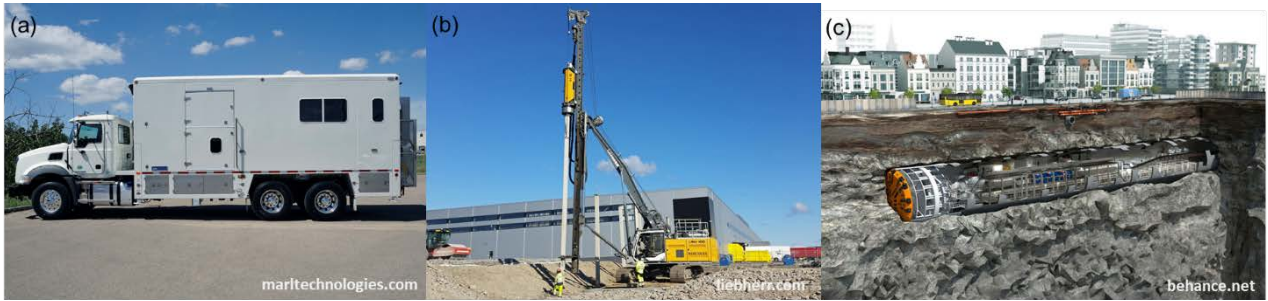

**Figure 1.** Geotechnical equipment for soil penetration and excavation: (a) rig for CPT and DMT testing, (b) hammer for pile driving, and (c) boring machine for tunnel construction.

Site investigation solutions have been developed to in part address the aforementioned challenges. For example, portable dynamic systems like the Dynamic Cone Penetrometer (DCP) (Escobar et al. 2016; Rollins et al. 2021) and geophysical testing equipment (Jamiolkowski 2012) can be more easily transported to the project site. However, despite the advances, these methods have limitations such as the shallow penetration depth for the former and the inability of providing various direct measurements of soil response and soil samples for the later.

Much work has been devoted to the development of so-called self-burrowing site investigation technology, which requires no external source of reaction force to penetrate the soil. Such equipment would have clear and far-reaching benefits in geotechnics if it could be transported to the site in small vehicles or carried by hand and if it provided the data necessary for design. These benefits span from potentially lowering the costs and carbon footprint of site characterization activities to increasing the productivity and allowing for characterization of previously inaccessible sites. In the future, untethered self-burrowing tools could enable unprecedented capabilities, such as underground steering with tight turning radii. While significant work is still required for the development of field-ready technology, important advances have been made towards this goal, particularly in the last five years.

**1.2. Bio-inspiration**

Animals, plants, and bacteria interact with soils in ways that are analogous to those between soils and human-made objects. Specifically, these organisms penetrate, excavate, and transfer load to soils, and these processes are controlled by the same physical phenomena as geotechnical applications, including site investigation (Martinez et al. 2022). Biological strategies have evolved through the process of natural selection; thus, they are efficient, multifunctional, and adaptable (Vogel 2000). Translation of strategies from the biological to the engineering domain can be done in terms of different levels of abstraction: forms, which consist primarily of physical structures, behaviors, which are specific movements and mechanisms, and principles, which are the underlying phenomena and processes that enable forms and behaviors to work (Fig. 2) (Mak and Shu 2004).

The field of bio-inspired geotechnics seeks to understand the geomechanical processes involved in biological strategies to adapt them towards geotechnical applications. This field has had a major focus on development of site investigation solutions, as described throughout this paper. Fig. 3 shows organisms that have been used as sources of inspiration for these advances, including earthworms, razor clams, plant roots, wasps, bees, fish and lizards. Work in other areas has developed piles and soil anchors inspired by snakeskin, soil anchors inspired by tree roots, drilling equipment inspired by angel wing shells and birds' beaks, and scour protection measures inspired by mangrove roots (Martinez et al. 2022; Martinez and Tao 2024).

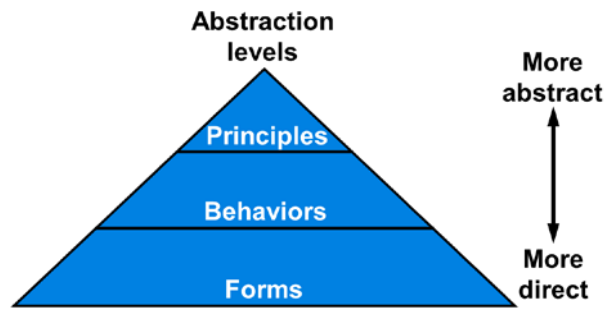

**Figure 2.** Levels of abstraction for translation of strategies from the biological to the engineering domain (adapted from Mak and Shu 2004).

**1.3. Paper scope**

This paper highlights research in bio-inspired site characterization that has been performed in the last decade towards the development of self-burrowing probes. Section 2 of this paper focuses on strategies to reduce the penetration resistance, which involve modifying the shape of the probe tip, expanding a shaft section near the tip, applying motions to the tip such as rotations and oscillations, and injecting fluids. In certain cases, the results show that the penetration resistance, and thus the required reaction force, can be decreased by a factor as high as 10 compared to direct static pushing. Section 3 centres on the analysis of self-burrowing probes, which through temporary anchors can generate the required reaction force to advance deeper into the soil. However, there are important interactions between the probe sections that control the probes' self-burrowing capabilities. The final section provides a review of the burrowing prototypes developed to date which and outlines research needs for transitioning this technology from the laboratory to the field.

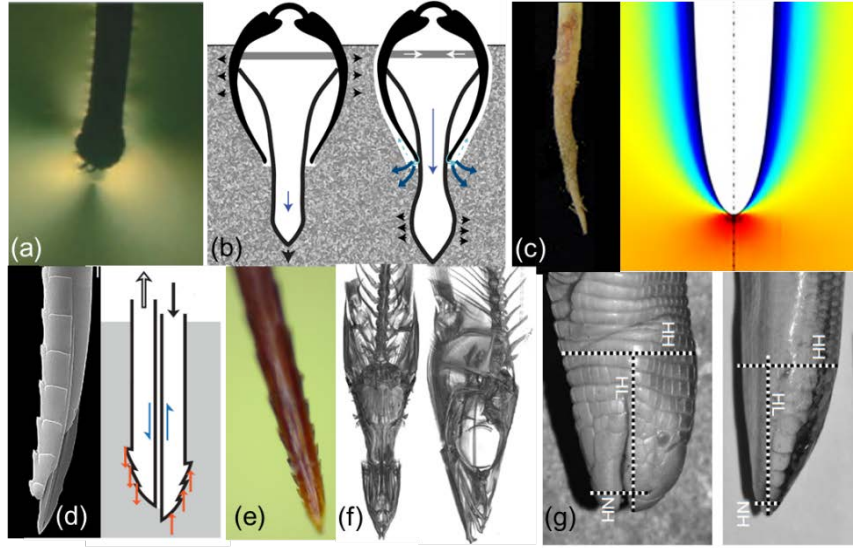

**Figure 3.** Biological adaptations for soil burrowing: (a) Marine worm in photoelastic material showing radial expansion of its tip and relaxation of effective stress ahead of tip (Dorgan et al. 2005), (b) schematic of dual anchor and local soil softening used by razor clam (Trueman 1968; Dorgan 2015), (c) photograph of root and simulation of radial growth of root relaxing stresses ahead of tip (Savioli et al. 2014), (d) SEM image of wasp ovipositor (Ghara et al. 2011) and reciprocal motion mechanism used to penetrate substrate (Cerkvenik et al. 2017), (e) honeybee stinger (Sahlabadi and Hutapea 2018), (f) top and side views of sand lance fish from CT Scan (Bizarro et al. 2016), and (g) photographs of lizard heads (Bergman and Berry 2021)

## 2. Reduction of penetration resistance

Invasive site investigation activities require inserting a probe or sampler into the ground, typically done by static pushing (e.g., CPT and DMT soundings), impact driving (e.g., SPT), and excavation (e.g., borings), all of which require a rig to provide the reaction forces. In contrast, animals and plants have developed strategies to penetrate soils without external sources of reaction force by addressing the high penetration resistance of soils ($q_c$) through various strategies. This section highlights recent work in geotechnics that has focused on decreasing the penetration resistance mobilized at both shallow and deep depths.

### 2.1. Tip shape

It has been long recognized that the shape of the tip of a probe or pile influences the resistance that is mobilized during penetration. Experience in the field has led to the understanding that flatter tips mobilize greater $q_c$ than sharper ones (Durgunoglu and Mitchell 1973; Lobo-Guerrero and Vallejo 2007; Tovar-Valencia et al. 2021). This is evident in the design of in-situ testing tools, where the standard CPT probe has a conical tip with an apex angle (α) of 60° and the standard DMT probe has a sharper blade-like tip to protect the probes and decrease the reaction force magnitude needed for insertion. In contrast, most piles have a flat tip to maximize the transfer of load at the base. This trend has also been recognized in the field of biology, indicating that "streamlining" has been developed in the heads of sand-diving lizards, stingers of honeybees and mosquitoes, and egg-laying organs of wasps to reduce the penetration resistance in soils and other substrates like wood (Kong and Wu 2009; Ling et al. 2016; Cerkvenik et al. 2017; Bergmann and Berry 2021).

The tip shapes that are most common in geotechnical engineering are either flat or conical, likely due to their symmetry and easy of machining. The earlier work by Durgunoglu and Mitchell (1973) and Koumoto and Houlsby (2001) provides analytical solutions indicating that $q_c$ increases with tip apex angle (α), tip surface roughness, and soil friction angle in both cohesive and cohesionless soils. While there is general agreement in these studies and later published data regarding the reduction in $q_c$ with α, the relationship between penetration resistance and tip shape depends on other factors such as depth, soil type, and soil density.

Hunt et al. (2023) performed penetration tests on sands using Discrete Element Modelling (DEM) simulations and centrifuge tests. The results indicate that for a given penetration depth, $q_c$ increases as α is increased (Fig. 4a-4d). This relationship can be quantified with a logistic equation with the following form, where the $q_c$ is normalized by that mobilized by a tip with an α of 60° ($q_{c,60°}$):

$$\frac{q_c}{q_{c,60°}} = A + \frac{\left(\frac{q_{c,max}}{q_{c,60°}} - \frac{q_{c,min}}{q_{c,60°}}\right)}{1+\left(\frac{\alpha}{I}\right)^{-k}} \quad (1)$$

where $q_{c,min}$ and $q_{c,max}$ are the maximum and minimum values of the relationship between $q_c$ and α, A is taken as the minimum $q_{c,i}/q_{c,60°}$ value, and k and I are fitting parameter that controls the rate of change and inflection point of the function, respectively. These results also indicate that the increase in $q_c$ diminishes as the depth is increased, with increases from a tip with an α of 30° to a flat tip ranging from 50% to 100% for a shallow penetration depth equivalent to 2 probe diameters ($Z/D_{probe}$ = 2) to increases between 25% and 50% at a depth of 5 probe diameters ($Z/D_{probe}$ = 5). In fact, greater increases in depth to a $Z/D_{probe}$ of 15 in the centrifuge tests results in negligible increases in $q_c$ with α. The results from other studies agree with these trends, with Fig. 4e showing an increase in $q_c$ with α and Fig. 4f showing a reduced increase with α as the penetration depth is increased.

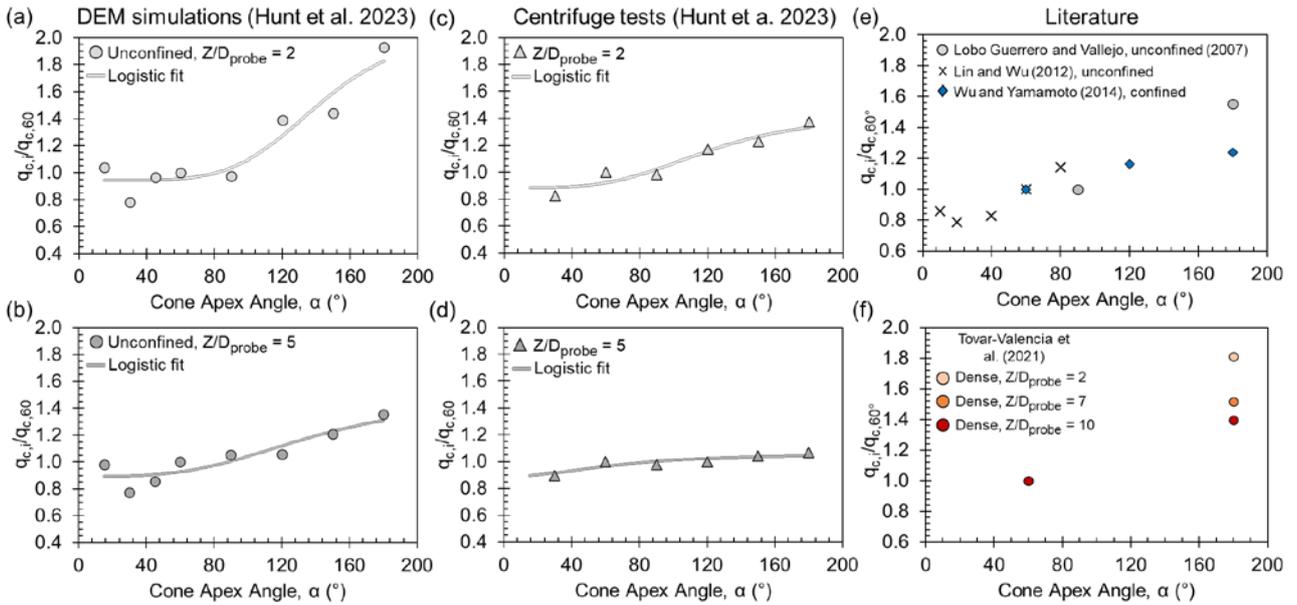

**Figure 4.** Relationship between normalized penetration resistance and cone apex angle from (a) and (b) DEM simulations, (c) and (d) centrifuge tests, and (e) and (f) literature (data from Hunt et al. 2023).

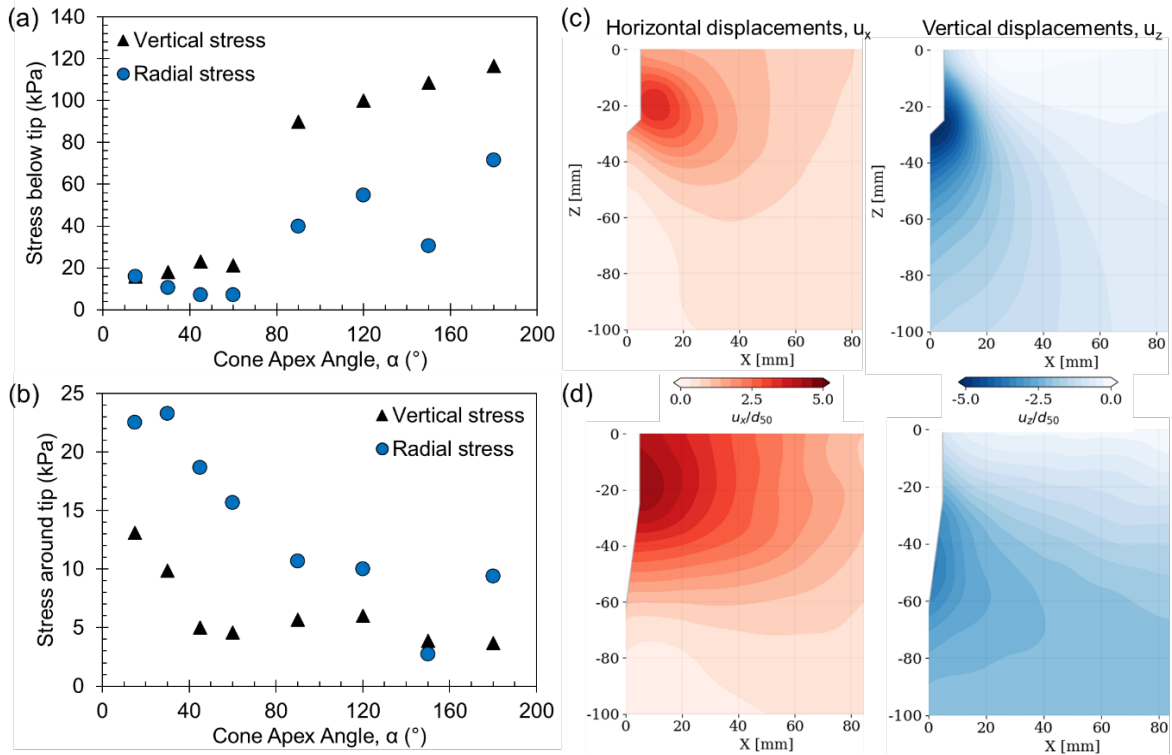

**Figure 5.** Meso-scale measurements of effective stresses (a) directly below and (b) laterally around penetrating probes with varying cone apex angle (data from Hunt et al. 2023) and particle displacement fields around a probe with an apex angle of (c) 90° and (d) 15° (data form Borela et al. 2021).

DEM simulation results shed light on the mechanisms leading to the change in $q_c$ with α. Local meso-scale measurements of vertical and radial effective stresses and particle displacements from Hunt et al. (2023) and Borela et al. (2021) show that greater apex angles increase soil stress magnitudes more and induce greater vertical displacements at locations directly below the tip (Fig. 5a and 5c), while smaller apex angles cause greater increases laterally around the tip accompanied by greater horizontal displacements (Fig. 5b and 5d). These results indicate that the blunter tips push soil down as they penetrate like a rigid punch and $q_c$ is largely a result of normal stresses acting on the probe tip. In contrast, the sharper tips push the soil radially away from the tip like an advancing wedge, where the $q_c$ originates mostly from frictional resistances at the soil-tip interface.

## 2.2. Expansion of a shaft section near the tip

High penetration resistances are a problem for the sprouting seeds of plants because the weight of the seed is the only reaction force available for growing roots. If the former is greater, the seed is lifted from the ground or the root is unable to penetrate the soil. To overcome this issue, roots have been observed to grow radially near the tip when they encounter a stiff soil layer, a process that has been shown to result in a reduction of the penetration resistance (Fig. 3c) (Barley 1962; Wilson et al. 1977; Bengough et al. 2011). Several species of earthworms and polychaetes (i.e., marine worms) have been observed to use a similar behavior, where a section of the worm close to the tip is expanded (Fig. 3a). In cohesive soils, this expansion can open a crack at the tip of the burrow due to tensile stress concentration, while in cohesionless soils it causes a relaxation of effective stresses (Dorgan et al. 2007; Dorgan 2015). In both instances, this reduces the strength of the soil at the burrow tip, causing a reduction in penetration resistance.

Research in geotechnical engineering has exploited this mechanism of expansion-aided reduction in penetration resistance. Chen et al. (2021) conceptualized a probe that consists of an anchor that can be radially expanded to a given expansion magnitude (EM) with a length of L and located a distance H behind the conical tip (Fig. 6a). Self-burrowing DEM simulations were performed on this probe in a virtual calibration chamber at a vertical effective stress of 100 kPa, showing reductions in $q_c$ when the anchor is expanded. Fig. 6b shows the time history of $q_c$ during initial insertion of the probe in a sand specimen (termed the CPT stage) followed by expansion of the anchor (termed the AE stage). In the figure, the time is normalized such that the CPT stage takes place from 0 to 1 and the AE stage takes place from 1 to 2. This reduction in $q_c$ is greatest for probes with anchors closer to the tip (i.e., smaller H), greater length, and greater expansion magnitude, as shown in Fig. 6c. The simulation results provide further evidence of the state of stresses around the probe, where the maps of particle contact forces show stress concentrations around the probe tip and anchor at the end of the AE stage (Fig. 7a). The difference in mean effective stresses (p') between the CPT and AE stages better highlights the effect of the anchor expansion, with decreases greater than 250 kPa at locations around the probe tip and increases greater than 250 kPa radially around the anchor (Fig. 7b). Similar reductions in $q_c$ as a shaft section is expanded have been reported in other studies, including Savioli et al. (2014), Ma et al. (2020), and Huang and Tao 2020).

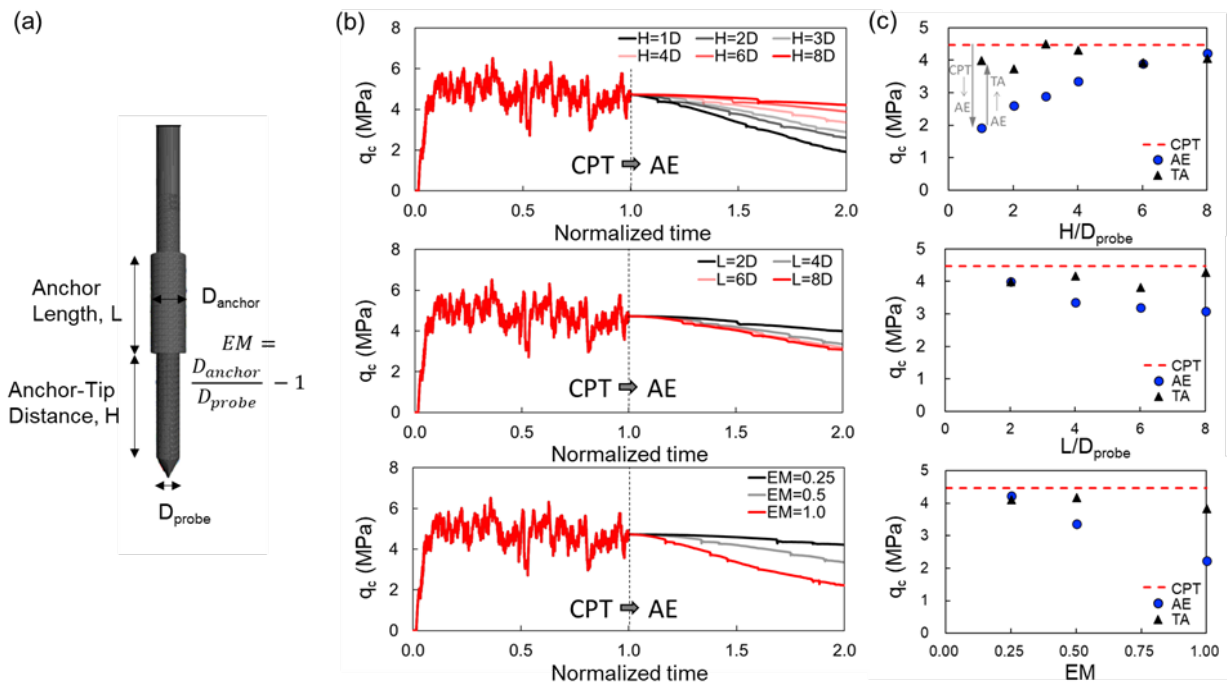

**Figure 6.** (a) Schematic of the self-burrowing probe simulated in DEM, (b) time history of penetration resistance during CPT advancement and anchor expansion, and (c) changes in penetration resistance due to anchor expansion and subsequent tip advancement (data from Chen et al. 2024).

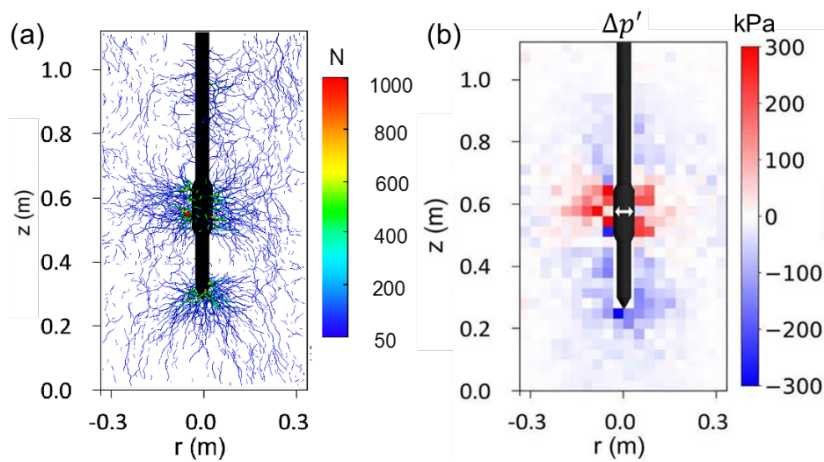

**Figure 7.** State of stress around a self-burrowing probe at the end of anchor expansion: (a) map of particle contact force magnitudes and (b) difference in mean effective stresses between the end of the cone penetration and anchor expansion stages, showing the decrease in stresses around the probe tip (data from Chen et al. 2022).

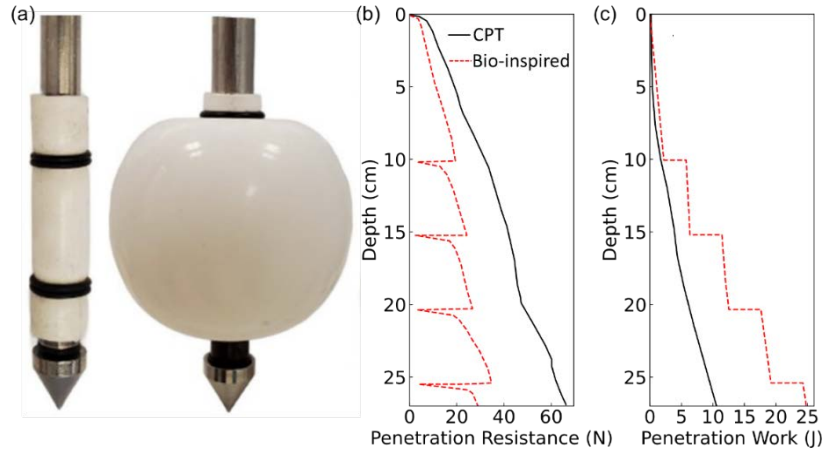

**Figure 8.** (a) Prototype of a probe with an expanding shaft section behind the tip (adapted from Naziri et al. 2024) and comparison of (b) penetration resistance and (c) work between soundings performed with the bio-inspired probe and a CPT probe (data from Naziri et al. 2024).

A third stage is performed in these simulations where the tip is advanced after the anchor is expanded (termed the TA stage). When the tip is advanced to distances greater than about 0.1 m, the $q_c$ is fully remobilized to the same magnitude measured during the initial CPT stage (Fig. 6c). This indicates that the reduction in tip resistance is temporary, but also shows that with sufficient tip advancement the CPT $q_c$ value can be obtained, thus allowing for use of existing methods for estimation of soil properties and geotechnical design (e.g. Mayne 2014; Lehane et al. 2020).

A prototype that uses the principle of expansion of a shaft section near the tip has been developed by Naziri et al. (2024), which provides experimental validation of the reduction in $q_c$. The prototype has a membrane immediately behind a conical tip, which can be expanded significantly, as shown in Fig. 8a. Experimental results of shallow soundings in sand show a reduction of about 50% in $q_c$ in comparison to a control CPT test (Fig. 8b), where the sharp drops take place during expansion of the membrane. After subsequent penetration, the $q_c$ magnitudes do not mobilize to values close to those during the CPT test, likely due to the large expansion magnitude of the membrane. The work involved in the penetration process indicates that the bio-inspired probe uses more energy due to the expansion of the membrane (Fig. 8c). In contrast, Huang and Tao (2020) performed DEM simulations of a probe with a similar configuration but smaller expansion magnitude and found a slight decrease in the total work with respect to a CPT test. These results show that expansion of a probe section behind the tip can be used to decrease the penetration resistance, and thus the required reaction mass of the equipment. However, depending on the magnitude of shaft expansion, the energy used may increase.

## 2.3. Tip oscillations

In addition to radial expansion of a region near the burrow tip, marine worms (i.e., polychaetes) have been observed to oscillate their heads from side to side. This strategy has been hypothesized to also help reduce the penetration resistance by pushing the soil laterally away from the burrow tip (Dorgan 2018). This strategy was implemented by Zhang et al. (2023) and Chen et al. (2024a) on DEM simulations of multi-cycle self-burrowing probe in sands of varying density at shallow depths. These simulations use the balance of vertical forces acting on the probe to determine whether it can self-burrow and advance deeper into the soil or it reaches refusal, as described in Section 3.3. Fig. 9a shows the particle normal contact forces acting on the probe tip during penetration with tip oscillations, showing the concentration of forces on the sides of the tip as it is being moved laterally. Comparison of the penetration resistances during penetration with oscillations to those mobilized during CPT penetration show significantly smaller values for the former. In fact, CPT penetration reached the penetration resistance refusal limit of 0.53 kN at a very small penetration distance of 0.06 cm (Fig. 9b), while the probe with tip oscillations continued to mobilize values well below 0.45 kN for greater penetration distances (Fig. 9c). In this data, the higher values correspond to stages when the tip coincided with the probe's longitudinal axis, while the lower values correspond to stages when the probe was fully extended laterally.

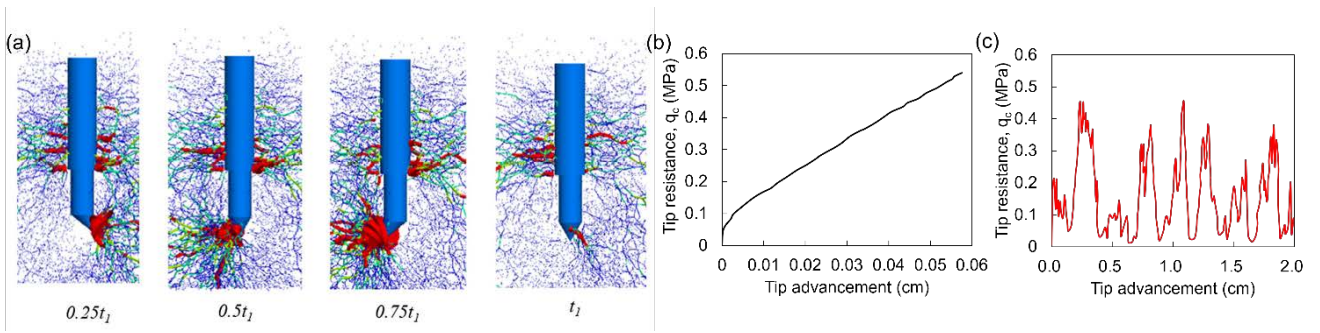

**Figure 9.** (a) Particle contact forces during tip oscillations and (b) comparison of penetration resistance during penetration with oscillations and a CPT sounding (data from Zhang et al. 2023).

## 2.4. Rotary motion

Several organisms have been observed to apply rotary motion while burrowing, such as the self-burying seeds of the common stork's bill plant (Evangelista et al. 2011; Jung et al. 2014). The seeds have awns that are sensitive to humidity, causing them to rotate under high humidity conditions. Similarly, the angled worm lizard actively rotates its head back and forth during burrowing (Gans 1968). In both instances, the rotary motion helps decrease

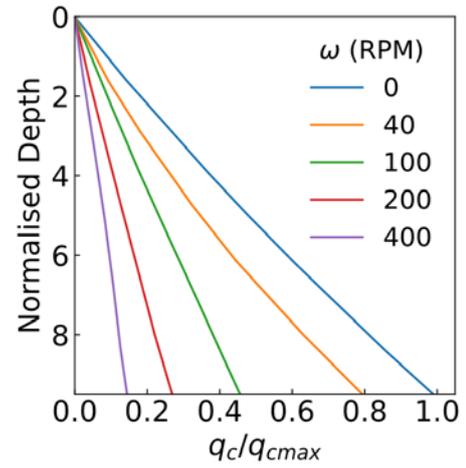

**Figure 10.** Simulation results of rotary penetration with varying angular velocities (data from Tang and Tao 2022).

the penetration resistance, allowing the organisms to further penetrate the soil.

Penetration simulations at shallow depths in DEM were performed by Tang and Tao (2022), showing the sharp decrease in $q_c$ as the angular velocity ($\omega$) is increased (Fig. 10). The authors show a reduction of over 80% for an $\omega$ of 400 rpm. Similar reductions in penetration resistance have been measured in geotechnically-focused studies, such as those involved in the rotary jacking of straight and screw piles (e.g., Sharif et al. 2021; Cerfontaine et al. 2023).

Another DEM study performed rotary CPT simulations in a virtual calibration chamber with an applied effective stress of 200 kPa (Yang et al. 2024). This study shows a similar decrease in $q_c$ with increased angular velocity, where the authors expressed it in terms of the ratio of the cone's tangential to vertical velocities (relative velocity = ωr/v, where r is the probe radius and v is the probe vertical velocity) (Fig. 11a). This was accompanied by an increase in the mobilized torque mobilized by the cone (Fig. 11b). The simulation results show that the decrease in $q_c$ is in part due to a reduction on the normal stress acting on the cone surface as the relative velocity is increased (Fig. 11c), which challenges previous analytical solutions (e.g., Bengough et al. 1997) and indicates that the rotary action decreases the strength of the soil located adjacent to the probe tip. Quantification of the work involved in the rotary penetration process (Fig. 11d) shows that it stays relatively constant at relative velocities smaller than about 1, while the $q_c$ decreases by about 15%. However, further increases in relative velocity which further decrease $q_c$ result in sharp increases in the total work done.

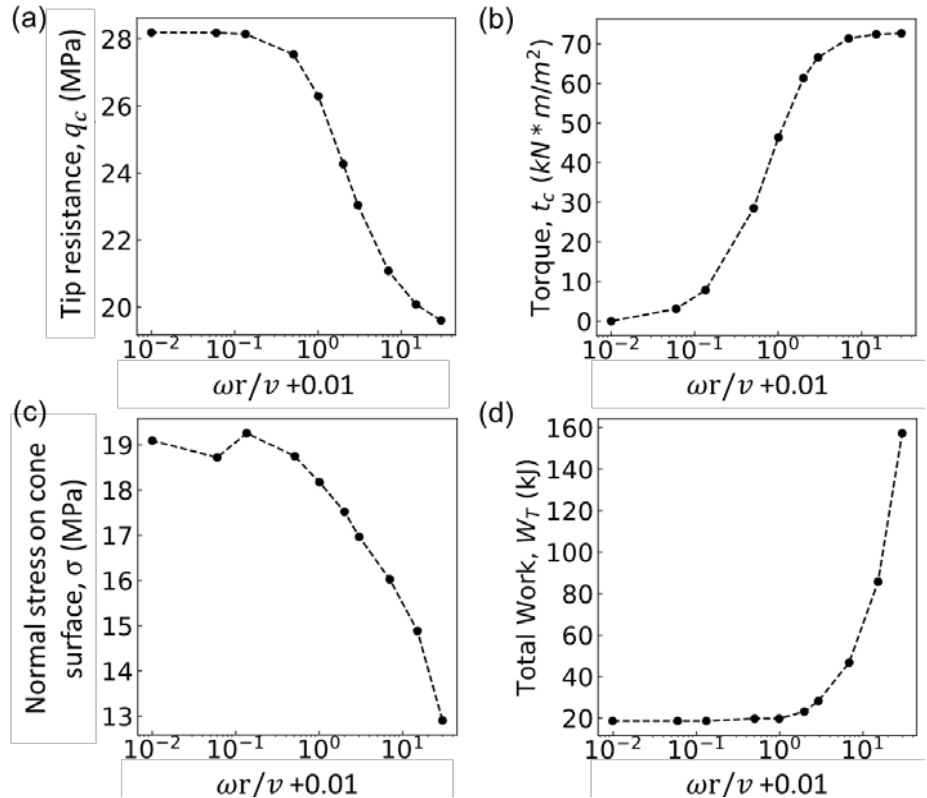

**Figure 11.** Results of DEM simulations of rotary penetration as a function of the cone relative velocity: (a) tip resistance, (b) torque, (c) stress acting on the cone surface, and (d) work done (data from Yang et al. 2024).

## 2.5. Root-inspired helical motion: circumnutations

Plant roots have developed a range of strategies to penetrate soils more efficiently. In addition to the ones described in Sections 2.2 and 2.4, the roots of many plants like rice, thale cress, pea, and maize have been observed to have a bent tip and grow into the soil following a helical path (e.g., Taylor et al. 2021; Simmons et al. 1995; Kim et al. 2016). This motion is referred to as circumnutations in the biological literature and has been shown to help in avoidance of obstacles (Taylor et al. 2021) and reduction of penetration resistance (Del Dottore et al. 2017). Design of a bio-inspired probe that can be used to investigate the effect of circumnutation-inspired motion (CIM) in soil penetration is shown in Fig. 12, showing the idealized design, experimental prototype, and paths of the probe tip with different relative velocities (defined above, i.e., ωr/v). Geometrical parameters of the probe include the bent angle (δ) of the tip and length of the bent portion ($L_1$).

Shallow penetration experiments performed with the probe shown in Fig. 12c attached to a robotic arm show sharp reductions in the penetration force ($F_z$) as the relative velocity is increased in sands of varying relative density and overconsolidated clay (Anilkumar and Martinez 2024; Anilkumar et al. 2024) (Figs. 13a and 13d). The probe used for these experiments had a δ of 10° and a $L_1$ equivalent to one probe diameter; however, the trends reported here been verified for different δ and $L_1$ values. As shown, at relative velocities greater than 1.5 π, the $F_z$ decrease to magnitudes that are less than 85% of those mobilized during a CPT sounding with a probe with the same diameter. The torque increases initially as the relative velocity is increased, then decreases or maintains constant values with additional increases (Figs. 13b and 13e). The calculated total work of penetration is shown in Figs. 13c and 13f, showing limited increases at small relative velocities followed by sharper increases, following a similar trend as shown for the rotary penetration in Fig. 11. These results show that a relative velocity of about 0.25π produces a decrease greater than 50% in the penetration force while only increasing the work done by about 25%. This could have important implications for site characterization, as the reduced $F_z$ would allow performing the sounding with smaller equipment whose mobilization to the site would have smaller environmental and economic impacts.

DEM simulations of shallow CIM penetration have shown a similar evolution of the penetration force, torque, and work with relative velocity (Chen and Martinez 2023). These simulations also show the differences in soil deformations as a result of CIM penetration in sands at shallow depths, where Fig. 14a-14c shows a spatial map where each particle's color is proportional to its displacement. CPT penetration produces a shallow failure in which a conical wedge of soil is lifted. The path of the probe tip is evident in the particles with large particle displacements for CIM penetration with 0.25π, while CIM penetration with 2π produces a more uniform mixing of the particles. The size of the soil wedge that is lifted during penetration decreases as the relative velocity is increased, suggesting that the soil disturbance is decreased. However, these trends should be verified at greater depths where the failure mechanism does not reach the soil surface. Particle contact normal forces are presented in Fig. 14d-

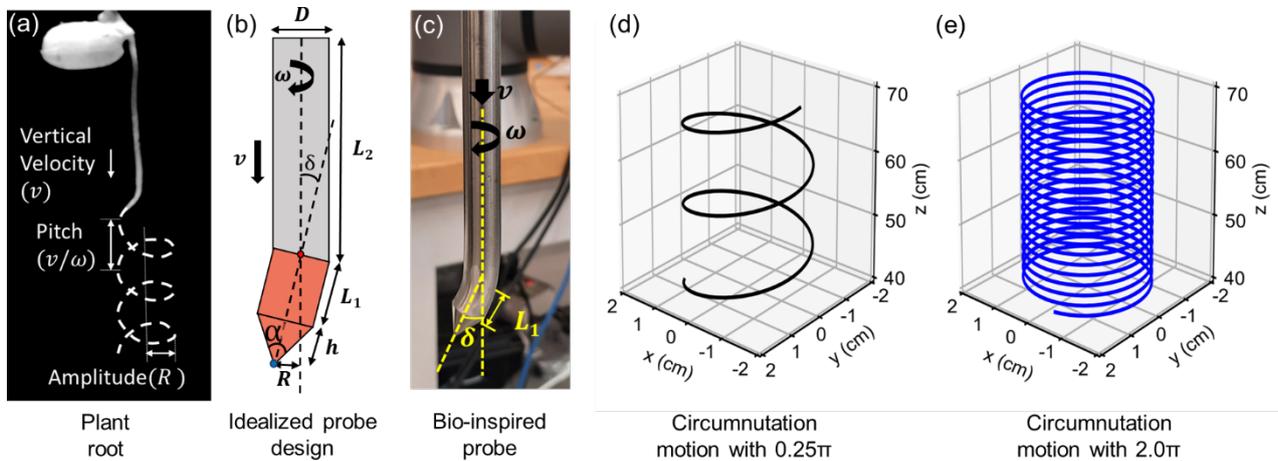

**Figure 12.** (a) Path of penetration of a plant root, (b) idealized design of a prototype probe that applies circumnutations-inspired motion, (c) photograph of a prototype probe, and paths of the probe tip with a (d) small and (e) large relative velocity.

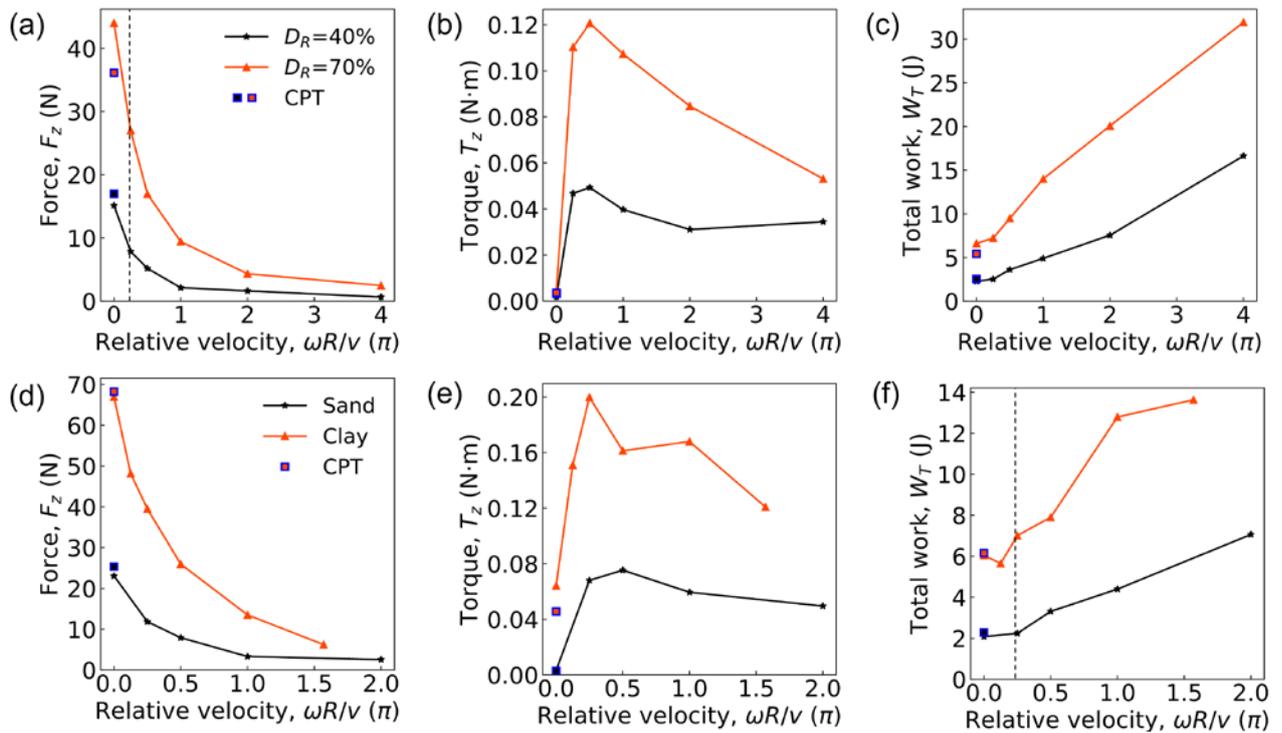

**Figure 13.** Mobilized vertical penetration force, torque, and total work for tests on (a) – (c) sand of varying relative density and (d) – (f) oversonsolidated clay and loose sand.

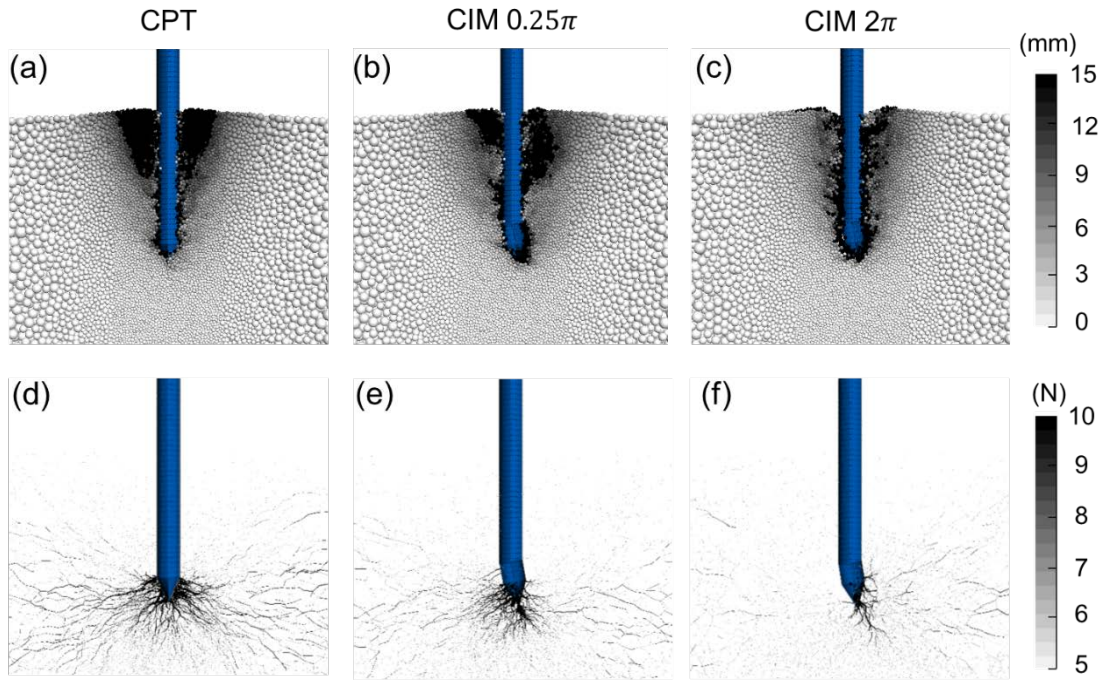

**Figure 14.** (a) – (c) Particle displacements and (d) – (f) particle contact forces during CPT and CIM penetration.

14f, showing a clear reduction in magnitude as the relative velocity is increased. The contact normal forces in the vicinity of the probe tip for the CIM $2\pi$ case are 80% smaller than those generated during CPT penetration (Chen and Martinez 2023).

**2.6. Soil fluidization by fluid injection**

Different organisms inject fluids in saturated soils to temporarily increase the pore pressures and thus reduce the penetration resistance. These include the southern sand octopus, Pacific sandfish, and several clam species that eject water jets from their body (Trueman 1967; Winter et al. 2012; MacDonald 2015; Montana et al. 2015). This strategy has also been implemented in geotechnical engineering to install piles (e.g., Passini et al. 2018; Passini and Schnaid 2015). With regards to self-burrowing probes, Naclerio et al. (2021) performed tests on a device that penetrates a dry sand deposit while it injects air at varying rates. As shown in Fig. 15, the penetration resistance decreases as the air flow rate is increased. However, the data shows a sharp increase in penetration force at a critical depth that increases with flow rate. While not explicitly discussed by the authors, this depth likely corresponds the overburden stress limit at which a given air flow rate is able to fluidize the sand.

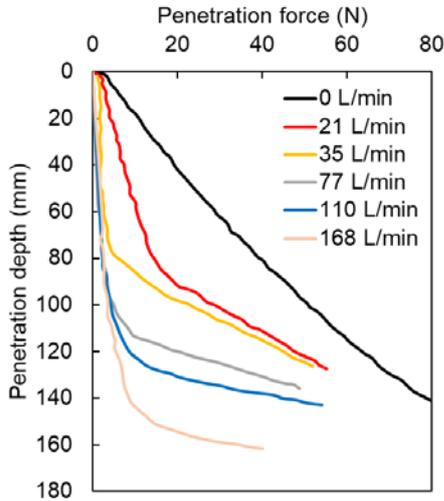

**Figure 15.** Results of penetration tests with varying flow rates of air injection (data from Naclerio et al. 2021).

## 3. Anchoring and self-burrowing

A probe's ability to successfully self-burrow hinges on the balance of resistance and reaction forces acting on it. The former typically consist of the penetration resistance at the tip and friction along the body of the probe, while the latter consist of both frictional and bearing forces acting along the probe's body mobilized largely by anchors that are temporarily deployed. This section summarizes research advances focused on the development of anchoring forces, interactions between the anchor and tip, and the assessment of the factors influencing the self-burrowing capabilities of probes.

### 3.1. Feasibility of self-burrowing

Analytical methods have been used to determine that probes can in theory self-burrow in a wide range of conditions relevant to geotechnical site characterization. In the simplest case, a self-burrowing probe can be idealized as having a cylindrical anchor that can be radially expanded to mobilize a limit radial pressure ($P_L$) and a conical tip that mobilizes a $q_c$ as its advanced, as shown schematically in Fig. 16a. The reaction force consists of friction on the anchor surface that depends on the $P_L$ magnitude, anchor's length (L), and expanded diameter ($D_e$), while the resistance force consists of the penetration resistance at the probe tip that depends on the $q_c$ magnitude and the probe diameter (D). A balance of vertical forces acting on the probe leads to the following equation, indicating the ratio of anchor length to probe diameter required to initiate self-burrowing:

$$\frac{L}{D} = \frac{q_c}{4(1+\varepsilon)\tau} \qquad (2)$$

where $\varepsilon$ is the expansion ratio of the probe (i.e., ($D_e$-D)/D, assumed as 20% in this work) and $\tau$ is the shear stress acting on the anchor surface. It is noted that this analysis ignores the bearing pressure mobilized at the top surface of the anchor as well as the friction mobilized along the probe shaft sections that are not expanded.

Martinez et al. (2020) performed cylindrical cavity expansion simulations using the ASCEND code (Jaeger 2019) which incorporates the MIT-S1 constitutive model (Pestana and Whittle 1999) to determine $P_L$ and $q_c$ values in clean sand, silty sand, silt, and clay. The simulation results on sand show an increase in $q_c/P_L$ ratio with a decrease in the state parameter (i.e., increase in density), which based on Eq. 2 lead to L/D ratios between 2.0 and 4.5 (Fig. 16b). The results in clay yield $q_c/P_L$ values that increase with overconsolidation ratio (OCR) and L/D between 3.0 and 4.2 (Fig. 16c). While the results for the silty sand and silt agree with those in clay and sand, they are not shown here for brevity. Interestingly, the L/D ratio of clam shells, which are the animals that most closely resemble the idealized probe geometry analysed by Martinez et al. (2020), ranges between 2.0 and 9.0, all of which are predicted to allow for self-burrowing based on the cavity expansion results. Overall, the cavity expansion analysis indicates that a probe with an anchor that has a length of about 5 times the probe base diameter and is expanded 20% can mobilize sufficient reaction forces to advance the tip forward in a wide range of soils and depths, which is supported by a review of calibration chamber and in-situ test results also presented by Martinez et al. (2020). However, there are important interactions between the probe's anchor and tip that influence the self-burrowing performance, as described in the following sections.

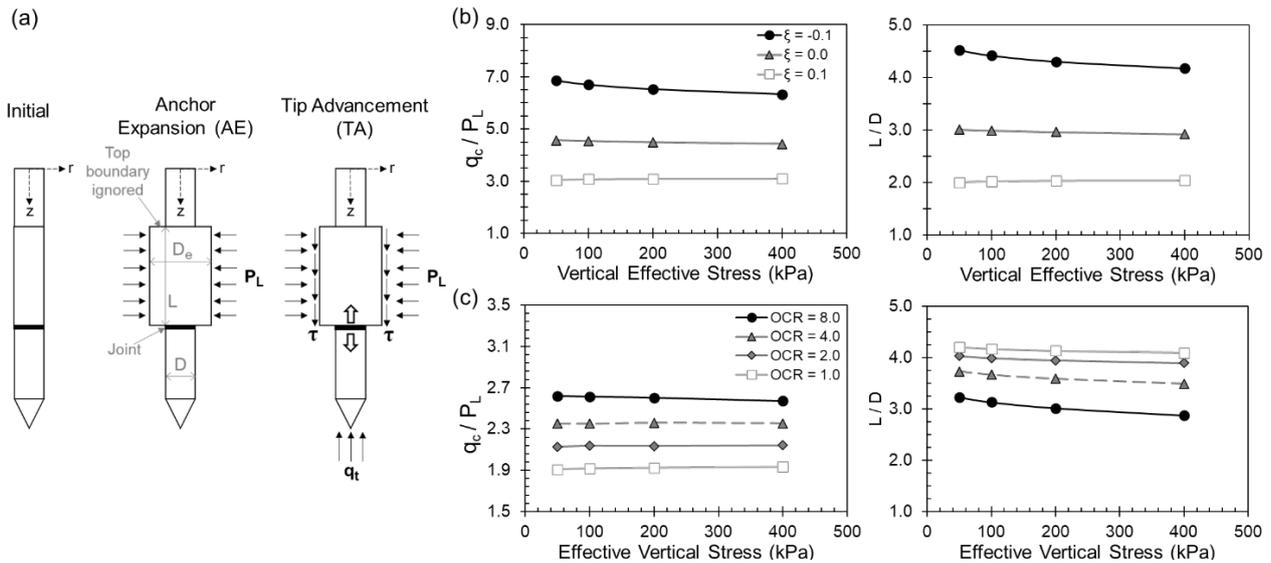

**Figure 16.** (a) Schematic of pressures acting on a self-burrowing probe analyzed using cavity expansion theory. Ratio of tip resistance to limit pressure and ratio of length to diameter of probe required for self penetration in (b) sands of varying initial state parameter and (c) clays of varying overconsolidation ratio (data from Martinez et al. 2020).

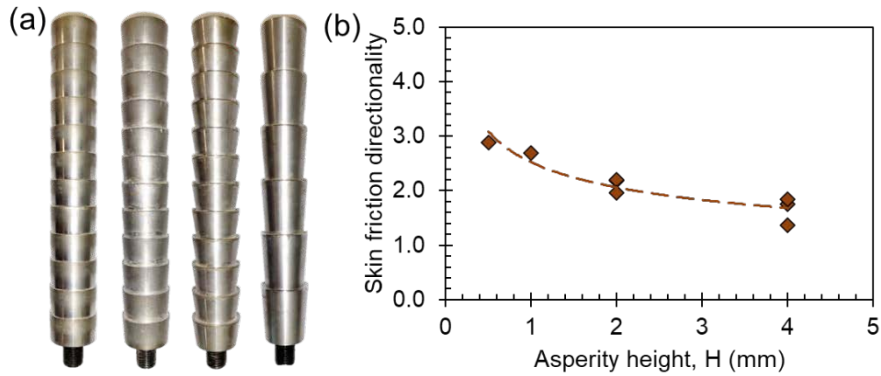

**Figure 17.** (a) Six-inch long sections of snakeskin-inspired soil anchors with different asperity heights and (b) ratio of skin friction mobilized during pullout to that mobilized insertion.

### 3.2. Direction-dependant friction

Direction-dependant friction, or frictional anisotropy, has been reported in many organisms to enable burrowing and locomotion. This strategy is enabled by the asymmetric shape of snake ventral (i.e., belly) scales (Gray and Lissmann 1950; Marvi and Hu 2012), small thorn-like structures within the awns of seeds (Kulić et al. 2009), and spike-like features in certain worms call setae (Merz and Edwards 1998). While direction-dependant friction has not been fully exploited with regards to self-burrowing technologies for site characterization, it would enable increased skin friction during anchorage and reduced friction resistances during advancement. Direction-dependant friction has already been applied towards soil-structure interfaces for applications in piles, soil anchors, and geosynthetics, showing its feasibility and robustness in a range of soil conditions, overburden stresses, and even in the field ( Martinez et al. 2019; O'Hara and Martinez 2022; Martinez et al. 2024; Gayathri and Vangla 2024). For example, Fig. 17 shows the ratio of skin friction mobilized during pullout to that during insertion of snakeskin-inspired soil anchors tested in the field by Martinez et al. (2024), which decreases with asperity height. As shown, the skin friction during pullout can be up to three times that during compression, highlighting the potential benefit to self-burrowing probes.

### 3.3. Tip advancement

Numerical simulations of self-burrowing probes consider the full complexity of the process, including the forces acting on all the sections of the probe and the interactions between them. Simulations have shown that in the same way that the penetration resistance decreases as the anchor of a self-burrowing probe is expanded, as previously described in Section 2.2 and shown in Figs. 6-8, the radial pressure on an expanded anchor decreases as it is loaded upwards (Chen et al. 2021; Chen et al. 2022; Huang and Tao 2020). These interactions are driven by arching within the soil and changes in principal stress directions.

Chen et al. (2021) and Chen et al. (2022) performed DEM simulations on the probe previously shown in Fig. 6a in a virtual calibration chamber under a vertical effective stress of 100 kPa. These simulations provided measurements of the evolution of the tip resistance and the radial pressure acting on the anchor's circumferential surface area ($P_a$) and the bearing pressure acting on the anchor's upper edge ($P_b$) (Fig.18). In the figure, the simulation time is normalized such that 0 to 1 indicates the initial penetration stage (CPT), 1 to 2 indicates the expansion of the anchor while the tip remains stationary (AE), and 2 to 3 indicates the tip advancement stage (TA) in which either the tip is moved downwards or the anchor is moved upwards based on a balance of vertical forces acting on the probe. During this stage, the section that mobilizes the smallest force moves. The results show the mobilization of $q_c$ during the CPT stage. In the AE stage, $q_c$ decreases gradually as the $P_a$ and $P_b$ pressures are mobilized during anchor expansion. In the TA stage, $P_a$ and $P_b$ remain constant at normalized times smaller than 2.5 during which the tip is advanced downward while the anchor remained stationary because the total reaction force was greater than the resistance force. A sudden drop in $P_a$ and an increase in $P_b$ take place at a normalized time of 2.5 when the anchor begins moving upward. At the end of the TA stage, the $q_c$ magnitude remobilizes to values close to those during the initial CPT stage.

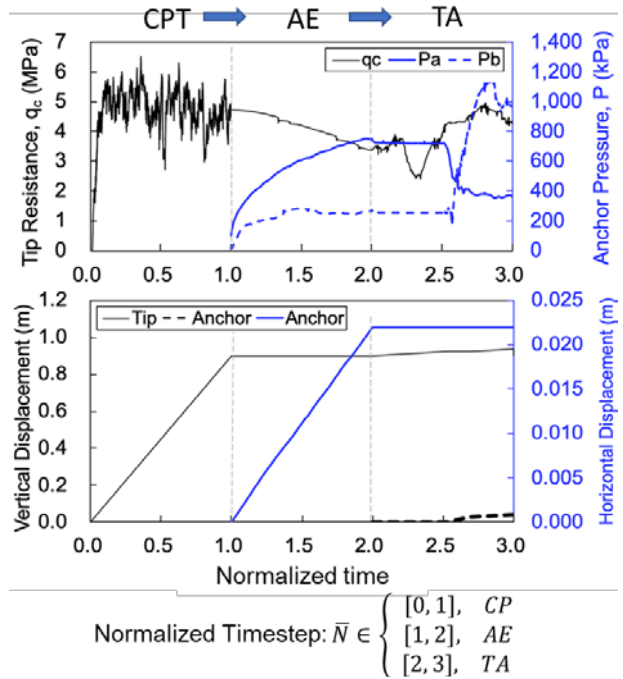

**Figure 18.** Self-burrowing DEM simulation results showing the evolution of tip resistance and anchor pressures during the initial insertion (CPT), anchor expansion (AE), and tip advancement stages (TA) (data from Chen et al. 2021).

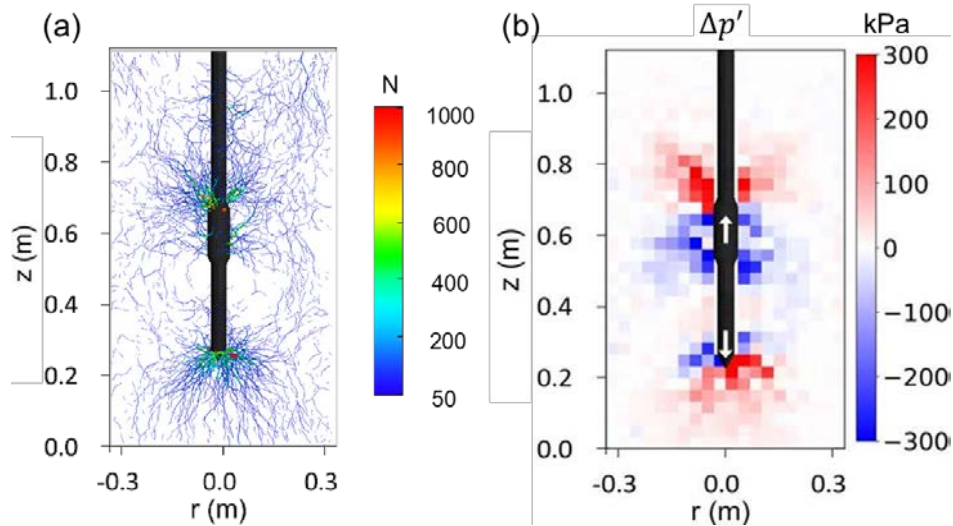

**Figure 19.** State of stress around a self-burrowing probe at the end of tip advancement: (a) map of interparticle contact force magnitudes and (b) difference in mean effective stresses between the end of the anchor expansion and tip advancement stages (data from Chen et al. 2022).

The trends in the pressures acting on the probe are explained by the reduction in particle contact force magnitudes at locations around the anchor and the increase in contact force magnitudes at locations behind the anchor (Fig. 19a). The changes in soil stresses can be better visualized by taking the difference in p' magnitude between the beginning and end of the TA stage. The spatial map of Δp' presented in Fig.19b clearly shows the relaxation that takes place around the anchor and the loading that takes place behind the anchor and below the tip. These changes in soil stresses are heavily influenced by the distance between the tip and anchor, anchor length, and anchor expansion magnitude (previously defined in Fig. 6a), which dictate the self-burrowing ability of the probe during the TA stage. The self-burrowing ability can be quantified by means of the self-burrowing distance, defined as the difference between tip and anchor displacements (ΔD). Positive ΔD values indicate net tip advancement to greater depths while negative values indicate net lifting of the probe (i.e., refusal conditions). The self-burrowing performance increases as the tip-anchor distance is decreased, anchor length is increased, and expansion magnitude is increased (Figs. 20a-20c).

In an effort to further increase the self-burrowing ability, Chen et al. (2024b) performed simulations on a probe with two anchors with varying spacing (S) (Fig. 21a). The simulation results show that the anchors influence each other, where the radial and bearing forces on the front anchor at the end of the TA stage ($F_{a2}$ and $F_{b2}$, respectively) decrease as the spacing is decreased (Fig. 21b-21c). However, the radial and bearing pressures on the back anchor ($F_{a1}$ and $F_{b1}$, respectively) are largely independent of S. These trends are explained by the particle displacement fields, showing significant overlap in the zones with large displacements around the closely-spaced anchors, while this overlap greatly reduces as the spacing is increased (Fig. 21d-21e). This influences the self-burrowing performance, with the self-burrowing distance increasing with inter-anchor spacing (Fig. 22a). Synthesis of the results from Chen et al. (2021) and Chen et al. (2024b) can be used to identify the probe configurations in terms of inter-anchor spacing, tip-anchor distance, and expansion magnitude that lead to successful self-burrowing from those that lead to refusal, as shown in Fig. 22b.

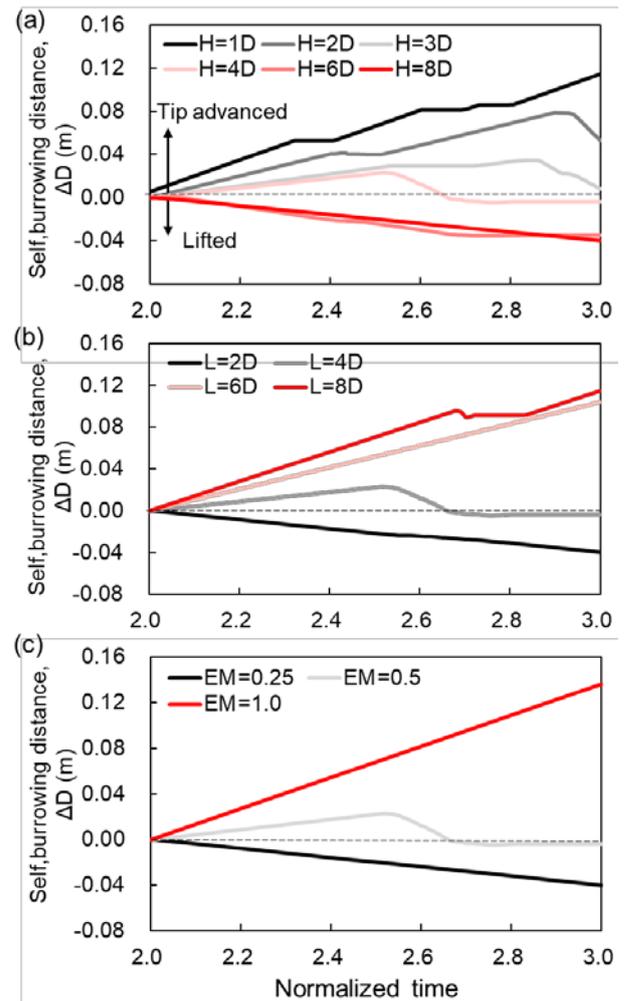

**Figure 20.** Self-burrowing distance during the TA stage for probes with varying (a) tip-anchor distance, (b) anchor length, and (c) anchor expansion magnitude (data from Chen et al. 2021).

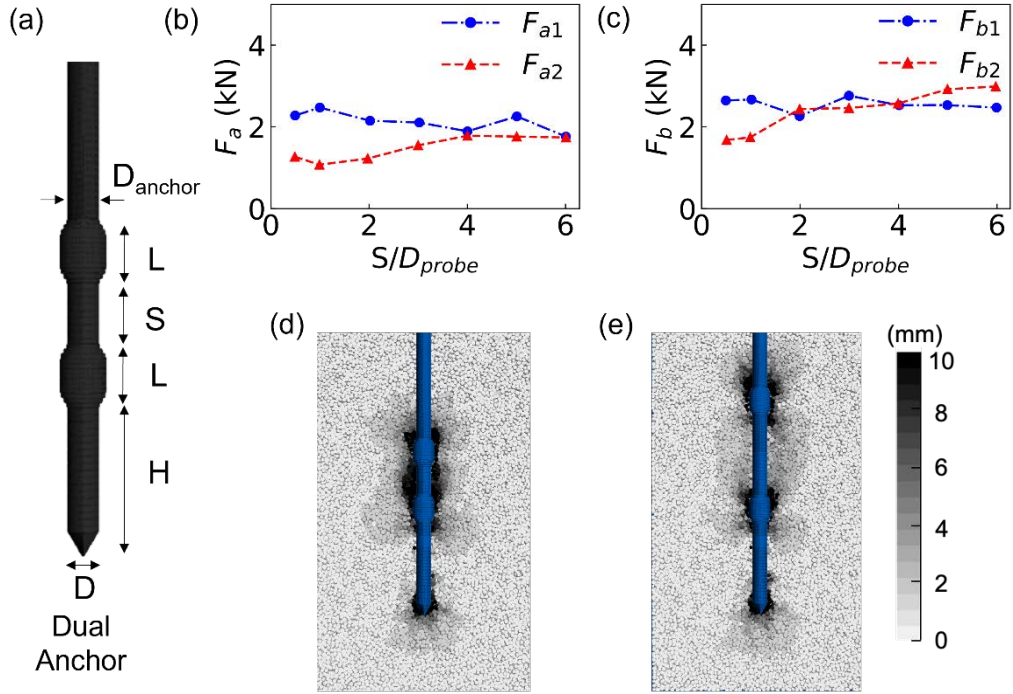

**Figure 21.** (a) Schematic of a probe with two anchors, (b) and (c) mobilized radial and bearing forces by the two anchors as a function of inter-anchor spacing, and spatial map of induced particle displacements with an S equivalent to (d) 2 and (e) 6 probe diameters (data from Chen et al. 2024b).

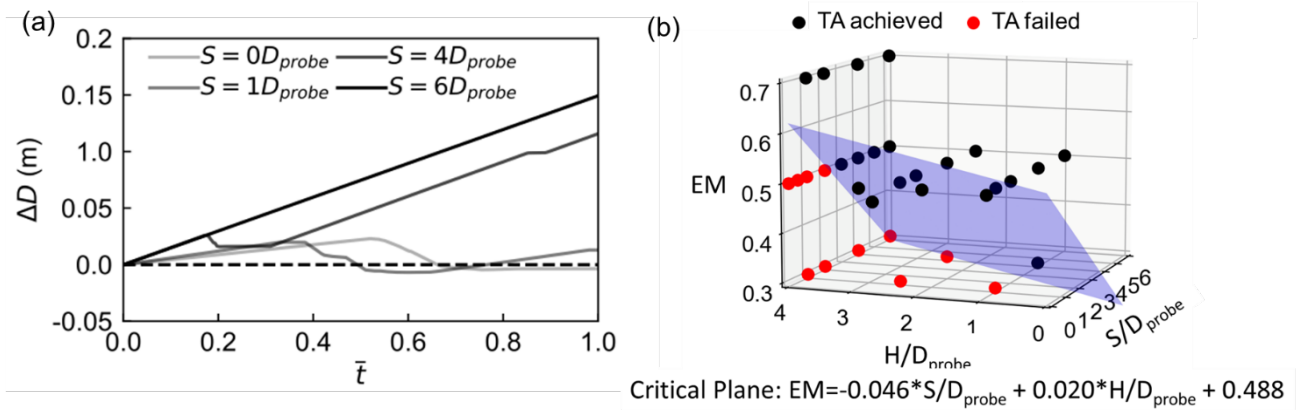

**Figure 22.** (a) Self-burrowing distance during the TA stage for probes with varying inter-anchor spacing and (b) tip advancement ability as a function of inter-anchor spacing, tip-anchor distance, and expansion magnitude (data from Chen et al. 2024b).

### 3.4. Multi-cycle self-burrowing

It is unlikely that an anchor that remains at a constant depth will provide sufficient reaction force to continue advancing the tip because the penetration resistance will in most cases continue increasing with depth. Therefore, a probe that can perform multiple self-burrowing cycles needs to have a minimum of two anchors. This motion is inspired by the dual-anchor burrowing strategy used by razor clams, which are highly efficient burrowers reaching depths beach sands as high as 70 cm in a few seconds (Trueman 1967). During burrowing, the razor clam first expands its shell to form a back anchor that allows the foot (i.e., tip) to penetrate the sand. Then, the foot is radially expanded to form a front anchor and the shell is contracted and then moved forward, allowing the clam to burrow deeper (Fig. 23a).

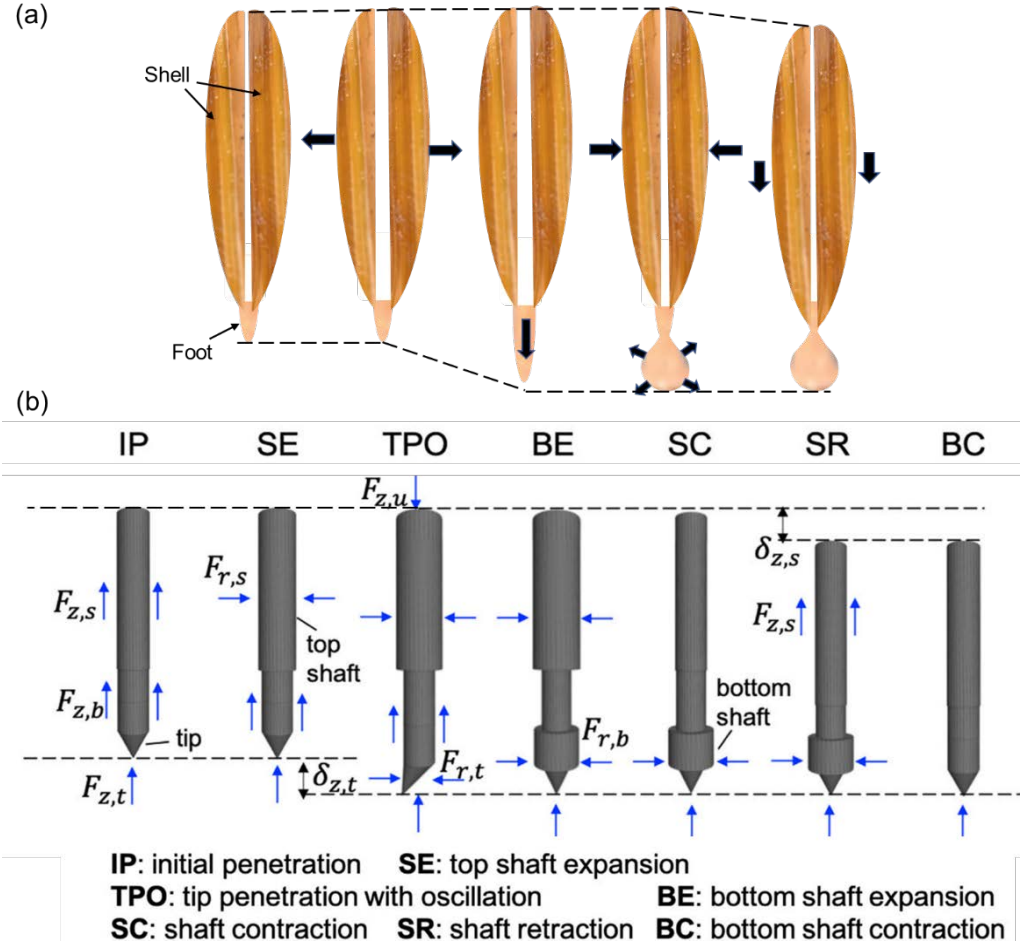

**Figure 23.** (a) Stages in the burrowing of a razor clam (adapted from Trueman 1967; Zhang et al. 2023) and (b) schematic of multi-cycle self-burrowing stages and associated forces acting on a bio-inspired probe (adapted from Chen et al. 2024a).

Zhang et al. (2023) and Chen et al. (2024a) performed shallow self-burrowing DEM simulations in sands of varying densities on a probe that implements a razor clam-inspired strategy, as shown schematically in Fig. 23b. The probe's motions are controlled by the balance of forces acting on it, such that it can only move to greater depths if the anchor reaction force is greater than the penetration resistance. The simulations demonstrated two important challenges for self-burrowing (i.e., net movement of the tip and anchor shown by a distance $\delta_{z,t}$, as shown in Fig. 23b): the decrease in radial pressure acting against the anchor once it is loaded and the high magnitude of penetration resistances. Self-burrowing was only possible if the shaft was continually expanded to maintain a constant radial pressure (to avoid the reduction previously shown in Figs. 18 and 19) and if the tip oscillated to decrease the penetration resistance (as shown in Figs. 9a and 9b).

Fig. 24a shows the evolution of forces acting on the probe, where $Q_z$ is the total penetration force, $F_z$ is the total reaction force, $F_{r,t}$ is the radial force acting on the anchor near the probe tip, and $F_{r,s}$ is the radial force acting on the back shaft anchor. As shown, the $Q_z$ magnitude remained significantly lower than $F_{r,s}$ due to the tip oscillations and the continued expansion of the back anchor. Once the tip is advanced a given distance (i.e., 3.5 cm in the first self-burrowing cycle and 7.5 cm in the second and third ones, Fig. 24b), the back anchor is contracted resulting in a drop in $F_{r,s}$, the anchor near the tip is expanded resulting in an increase in $F_{r,t}$, and then the back anchor is retracted forward. It is noted that in these simulations the relatively small self-burrowing distances (i.e., in order of cm) are limited by the computational cost of the simulations, rather than by the self-burrowing ability of the probe. Quantification of the work done during the different stages of the simulation show that oscillation of the tip accounts for about 60% of the work done by the end of the simulation, expansion the front and back anchors accounts for about 20% and 13%, respectively, and advancement of the tip downwards accounts for less than 10% (Fig. 24c) (Chen et al. 2024b). When compared to a CPT sounding in which all the work done is due to the penetration resistance and the friction along the shaft, the self-burrowing probe does about 25% more work (Zhang et al. 2023). Further investigation focused on fine-tuning the probe motions to enable self-burrowing while minimizing the work done could improve the energetic efficiency. Nonetheless, these results show that self-burrowing at shallow depths is possible from a geomechanical point of view with a small amount of additional work, which would likely be much smaller than the energy required in mobilizing a conventional rig to the project site.

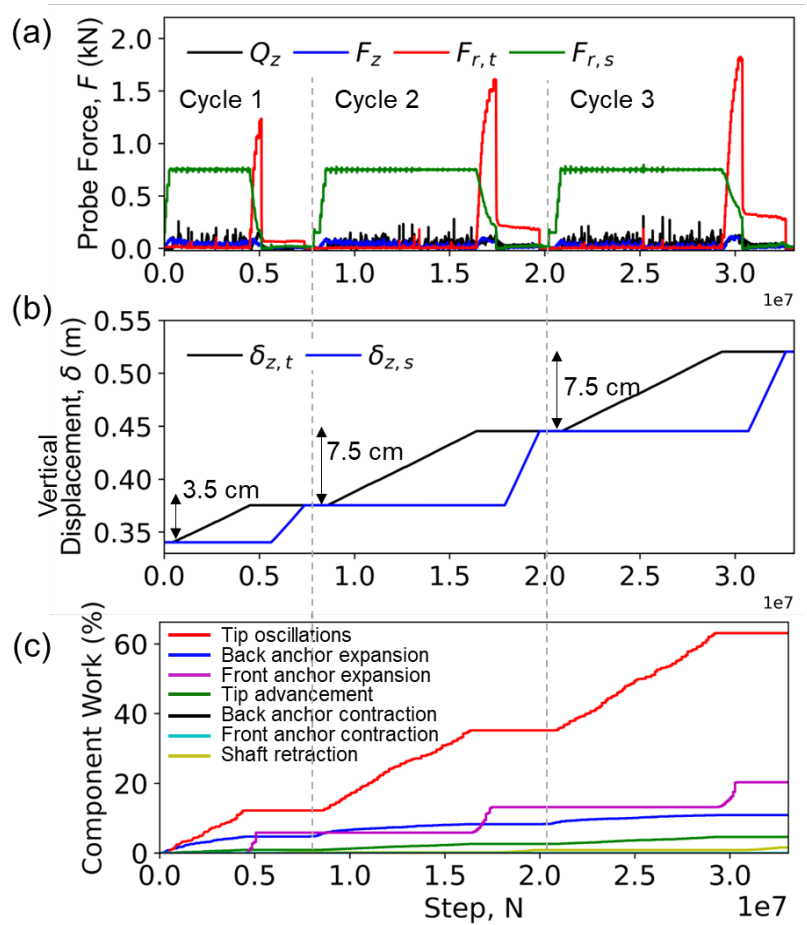

**Figure 24.** Evolution of (a) forces acting on probe, (b) tip and back shaft displacements, and (c) work done during a self-burrowing simulation with three cycles in medium-dense sand (data from Chen et al. 2024a).

## 4. Self-burrowing prototypes, challenges, and implications in practice

Several burrowing prototypes have been developed in the fields of geotechnics and robotics during the last decade. This section first provides a review of recent work to show the rapid advances that have been made to date. Finally, the paper is concluded with discussion of the research needs and challenges that need to be addressed to transfer the developments in self-burrowing technologies from the laboratory to the field.

### 4.1. Self-burrowing probe prototypes

Prototypes that fluidize soil have been shown to successfully burrow in the laboratory and the field. Winter et al. (2014) developed a razor-clam inspired device that is designed to fluidize the surrounding soil by dynamically cycling its foot up and down and shaft in and out (Fig. 25a). The device was tested in the field in a marine mudflat, in which it required an applied external force of about 180 N to burrow to a depth of 200 mm. When tested in the laboratory in glass beads, it was able to self-burrow (i.e., without an external force) to a depth of almost 300 mm. The test results indicate that the RoboClam uses between 80% and 900% more energy to reach the same depth than statically pushing an object with the same diameter. Naclerio et al. (2021) developed a burrowing robot that is inspired by plant roots which grow only at their tips, burial of octopi which fluidizes the sand at the seabed by a fluid jet, and asymmetry of the head of sandfish lizards to allow for steering (Fig. 25b). This prototype was tested in a dry clean sand and used air for fluidization, as previously described in Section 2.6 and Fig. 15. It was able to burrow vertically to depths up to 35 cm at a velocity of 2 cm/s as well as to burrow horizontally at a depth of 8 cm for 60 cm at a velocity of 2 cm/s. While both these prototypes successfully burrowed in soils, the reliance on fluidization by dynamic action or fluid injection likely limits their performance at greater depths and in cohesive soils.

Other prototypes have mostly been tested in laboratory conditions. Ortiz et al. (2019) developed a device that uses the dual anchor strategy combined with tip oscillations (Fig. 25c). The prototype was able to burrow horizontally in an idealized material composed of polypropylene pellets at a depth of 5 cm for a distance of 10 cm. Bagheri et al. (2024) developed one of the largest devices reported in the literature, with a diameter of 27 cm and length of 20 cm which uses tip rotation with an auger tip (Fig. 25d). The prototype was able to self-burrow downwards in dry glass beads for distances of about 200 mm at velocities between 12 and 1 mm/s, with greater rotation speeds leading to greater burrowing velocities. Zhong et al. (2023) built a cylindrical robot that combines the effects of tip shape and rotary motion to enable horizontal self-burrowing. The device is equipped either a conical or auger shape which is pushed forward by a linear actuator, while the back end of the device has a flat tip (Fig. 25e). The device burrowed for about 9 cm at a depth of 15 cm in dry glass beads.

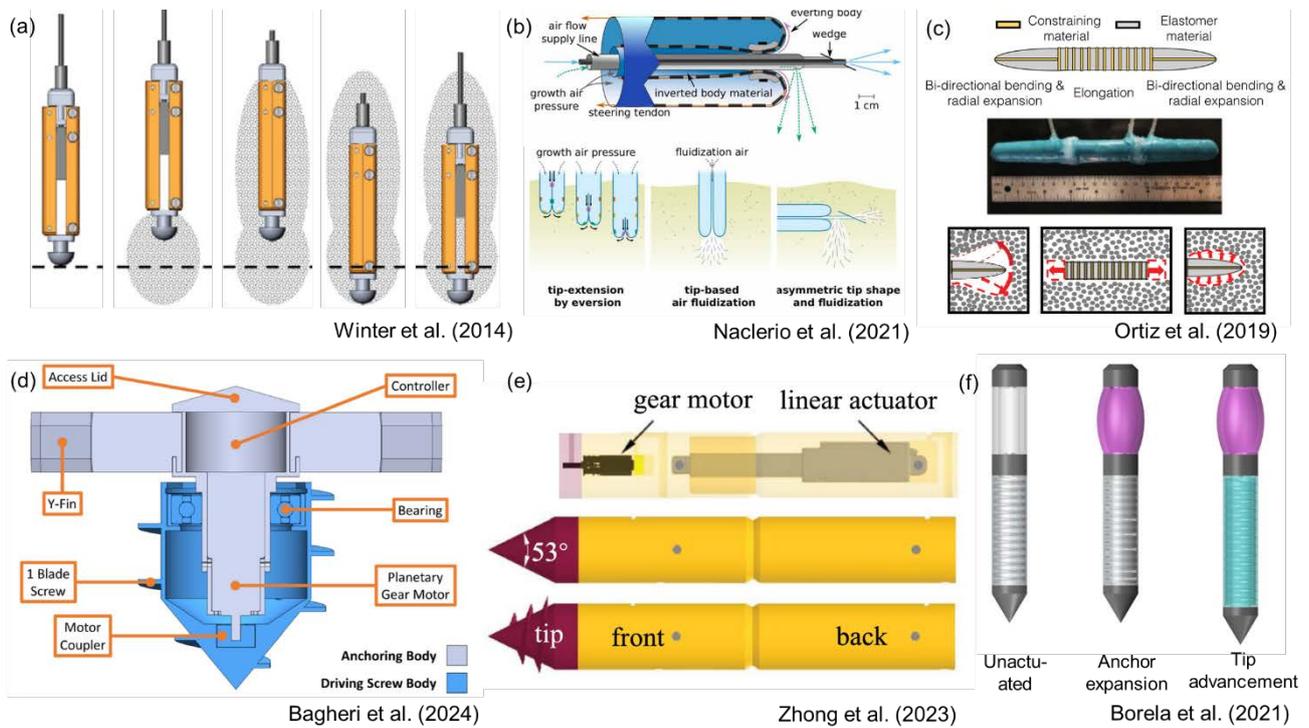

**Figure 25.** Self-burrowing prototypes that employ different mechanisms: (a) fluidization, (b) fluidization and tip eversion, (c) tip oscillations and radial expansion, (d) rotary motion, (e) rotary motion and extension, (f) tip and anchor.

Borela et al. (2021) developed a prototype that has one anchor that is radially expanded and a section that is elongated axially to push the tip forward (Fig. 25f), which most resembles the probe simulated in DEM (Fig. 6a). The device was buried to depths between 315 and 365 mm in dry Hostun sand and was only able to advance its tip by a maximum distance of 8 mm. The small tip advancement achieved by this prototype clearly highlights the challenge associated with the high penetration resistance magnitudes involved in static pushing. Indeed, it appears that to successfully self-burrow, a probe needs to either decrease the penetration resistance or use a larger source of reaction force, such as a rig or anchor(s) with large dimensions. The former strategy may be the most feasible for self-burrowing; however, it precludes obtaining a tip resistance value equivalent to the CPT $q_c$ for which established methods exist to estimate soil properties and perform geotechnical design. In this case, the probe could be used to install instrumentation or obtain samples, while new relationships for engineering use would need to be developed.

## 4.2. Perspectives on future research needs and challenges

This concluding section provides some perspectives on future research needs and challenges that need to be addressed to develop self-burrowing technology that is field-ready for geotechnical site investigation. Self-burrowing site investigation equipment or equipment that can be transported in light weight vehicles can bring significant benefits for geotechnical design, construction. However, while important advances have been made in the last decade towards this goal, Section 4.1 highlighted the significant gaps between the capabilities of the current prototypes and conditions that are relevant to geotechnical site characterization. The following aspects need to be addressed to transition the technologies to the field. In doing so, advances to the fundamental understanding in machine-soil interactions, soil behavior, and animal and plant locomotion will be achieved.

*Soil types:* With a few exceptions, most of the work to date has considered sandy soils, likely due to computational and experimental challenges. The mechanisms of penetration resistance reduction and anchorage should also be evaluated a broader range of natural soils, including clays, silts, and gravels.

*Limited depth*: While some of the numerical analyses have considered greater overburden pressures, all experiments performed to date have been at depths smaller than 0.5 m. Evaluation at greater depths is needed to fully characterize the efficiency and identify limitations of the bio-inspired strategies.

*Anchorage:* A large proportion of the work has focused on the reduction of penetration resistances. However, less emphasis has been placed on the mobilization of anchorage forces that can be maintained at a wide range of depths. This work would be largely geotechnical in nature, focusing on load transfer and soil failure mechanisms.

*Energy efficiency:* While reducing the size or need of mobilizing a rig to the site would greatly reduce the energy spent, the efficiency of the bio-inspired penetration processes in isolation is typically smaller than static pushing. Research that fine tunes the bio-inspired processes can lead to increased efficiency in addition to successful self-burrowing.

*Estimation of soil properties and use in engineering design*: It is likely that successful self-burrowing will require use of processes that reduce the magnitude of penetration resistance, eliminating the readings of established in-situ tests (e.g., CPT $q_c$, $f_s$, and $u_2$) for which design methods have been calibrated. Research should be undertaken to develop correlations between readings with bio-inspired penetration strategies (i.e., with circumnutations or tip oscillations) and established in-situ test readings. Alternatively, the use of self-burrowing technologies could be limited to the installation of instrumentation or retrieval of soil samples.

*Integration with site investigation equipment*: Once the technology has been developed and the capabilities proven, the new tools need to be integrated with existing site investigation equipment and components (e.g., electronics, drill rods, hydraulic actuators) to facilitate adoption by industry. In addition, standards need to be developed to ensure accuracy and reproducibility.

*Cost and practicality*: The new tools need to be economically competitive in comparison to established in-situ testing tools, and the processes involved in the performance of the tests need to be practical to ensure appropriate productivity in the field.

*Soil proofing*: All of the bio-inspired strategies involve moving parts to reduce the penetration resistance or generate anchorage. Keeping soil out of hinges, connections, and points of relative motion will necessitate design of dynamic seals to ensure durability and continued use of the equipment. It is likely that this research will necessitate collaboration between mechanical and geotechnical engineers.

*Communication and power*: Self-burrowing tools can either be tethered or untethered. The former is similar to the state of practice, where both the data and power are transmitted through cables. The later opens a host of new capabilities, such as underground steering, but it involves significant challenges in the transmission of data through the soil and power provided by batteries for soundings that are sufficiently deep. These advances will likely be in the fields of electrical and mechanical engineering.

## Acknowledgements

The interactions and discussion with several individuals, including Jason DeJong from UC Davis, David Frost from Georgia Tech, and Julian Tao from Arizona State University, and the broader community of the Center for Biomediated and Bioinspired Geotechnics (CBBG), have been of great value in the development of the material presented in this paper. Much of the material presented in based on the work of graduate students and postdocs, including Olivia Hunt, Kyle O'Hara, Damon Nguyen, Hyeon Jung Kim, Lin Huang, Mandeep Basson, Aly Ghanem, Ali Khosravi, Ningning Zhang, Xiaotong Yang, and Bowen Wang. The first author would like to acknowledge the collaboration with Raul Fuentes from RWTH Aachen and Rui Wang from Tsinghua University. This material is based upon work supported by the Engineering Research Center Program of the National Science Foundation (NSF) under NSF Cooperative Agreement No. EEC–1449501 which funds the CBBG and by the NSF under Award No. 1942369. Any opinions, findings, and conclusions or recommendations expressed in this material are those of the author(s) and do not necessarily reflect those of the National Science Foundation.

## References

Anilkumar, Riya, Yuyan Chen, and Alejandro Martinez. 2024. "Plant Root-Inspired Soil Penetration in Sands Using Circumnutations for Geotechnical Site Characterization." In *Geo-Congress 2024*, 252–62. Vancouver, British Columbia, Canada: American Society of Civil Engineers. https://doi.org/10.1061/9780784485309.026.

Anilkumar, Riya, and Alejandro Martinez. 2024. "Plant Root Circumnutation-Inspired Penetration in Sand and Clay." In *Proceedings of the 7th International Conference on Geotechnical and Geophysical Site Characterization*. Barcelona, Spain: ISSMGE Technical Committee TC 102.

Bagheri, Hosain, Daniel Stockwell, Benjamin Bethke, Nana Kwame Okwae, Daniel Aukes, Junliang Tao, and Hamid Marvi. 2024. "A Bio-Inspired Helically Driven Self-Burrowing Robot." *Acta Geotechnica* 19 (3): 1435–48. https://doi.org/10.1007/s11440-023-01882-9.

Barley, K. P. 1962. "The Effects of Mechanical Stress on the Growth of Roots." *Journal of Experimental Botany* 13 (1): 95–110. https://doi.org/10.1093/jxb/13.1.95.

Bengough, A. G., C. E. Mullins, and G. Wilson. 1997. "Estimating Soil Frictional Resistance to Metal Probes and Its Relevance to the Penetration of Soil by Roots." *European Journal of Soil Science* 48 (4): 603–12. https://doi.org/10.1111/j.1365-2389.1997.tb00560.x.

Bengough, A. Glyn, B. M. McKenzie, P. D. Hallett, and T. A. Valentine. 2011. "Root Elongation, Water Stress, and Mechanical Impedance: A Review of Limiting Stresses and Beneficial Root Tip Traits." *Journal of Experimental Botany* 62 (1): 59–68. https://doi.org/10.1093/jxb/erq350.

Bergmann, Philip J., and David S. Berry. 2021. "How Head Shape and Substrate Particle Size Affect Fossorial Locomotion in Lizards." *Journal of Experimental Biology* 224 (11): jeb242244. https://doi.org/10.1242/jeb.242244.

Bizarro, Joseph J., A. N. Peterson, J. M., Blaine, J.P. Balaban, H.G. Greene, A.P. Summers. 2016. "Burrowing behavior, habitat, and functional morphology of the Pacific sand lance (Ammodytes personatus)." Fishery Bulletin, 114(4): 445-460.

Borela, R., J. D. Frost, G. Viggiani, and F. Anselmucci. 2021. "Earthworm-Inspired Robotic Locomotion in Sand: An Experimental Study Using x-Ray Tomography." *Géotechnique Letters* 11 (1): 1–22. https://doi.org/10.1680/jgele.20.00085.

Cerfontaine, Benjamin, Michael John Brown, Jonathan Adam Knappett, Craig Davidson, Yaseen Umar Sharif, Marco Huisman, Marius Ottolini, and Jonathan David Ball. 2023. "Control of Screw Pile Installation to Optimise Performance for Offshore Energy Applications." *Géotechnique* 73 (3): 234–49. https://doi.org/10.1680/jgeot.21.00118.

Cerkvenik, Uroš, Bram Van De Straat, Sander W. S. Gussekloo, and Johan L. Van Leeuwen. 2017. "Mechanisms of Ovipositor Insertion and Steering of a Parasitic Wasp." *Proceedings of the National Academy of

*Sciences* 114 (37). https://doi.org/10.1073/pnas.1706162114.

Chen, Yuyan, Ali Khosravi, Alejandro Martinez, and Jason DeJong. 2021. "Modeling the Self-Penetration Process of a Bio-Inspired Probe in Granular Soils." *Bioinspiration & Biomimetics* 16 (4): 046012. https://doi.org/10.1088/1748-3190/abf46e.

Chen, Yuyan, and Alejandro Martinez. 2023. "DEM Modelling of Root Circumnutation-Inspired Penetration in Shallow Granular Materials." *Géotechnique*, September, 1–18. https://doi.org/10.1680/jgeot.22.00258.

Chen, Yuyan, Alejandro Martinez, and Jason DeJong. 2022. "DEM Study of the Alteration of the Stress State in Granular Media around a Bio-Inspired Probe." *Canadian Geotechnical Journal* 59 (10): 1691–1711. https://doi.org/10.1139/cgj-2021-0260.

Chen, Yuyan, Ningning Zhang, Raul Fuentes, and Alejandro Martinez. 2024a. "A Numerical Study on the Multi-Cycle Self-Burrowing of a Dual-Anchor Probe in Shallow Coarse-Grained Soils of Varying Density." *Acta Geotechnica* 19 (3): 1231–50. https://doi.org/10.1007/s11440-023-02088-9.

Chen, Yuyan, Alejandro Martinez, and Jason DeJong. 2024b. "DEM Simulations of a Bio-Inspired Site Characterization Probe with Two Anchors." *Acta Geotechnica* 19 (3): 1495–1515. https://doi.org/10.1007/s11440-022-01684-5.

Del Dottore, Emanuela, Alessio Mondini, Ali Sadeghi, Virgilio Mattoli, and Barbara Mazzolai. 2017. "An Efficient Soil Penetration Strategy for Explorative Robots Inspired by Plant Root Circumnutation Movements." *Bioinspiration & Biomimetics* 13 (1): 015003. https://doi.org/10.1088/1748-3190/aa9998.

Dorgan, Kelly M. 2015. "The Biomechanics of Burrowing and Boring." *Journal of Experimental Biology* 218 (2): 176–83. https://doi.org/10.1242/jeb.086983.

Dorgan, Kelly M.. 2018. "Kinematics of Burrowing by Peristalsis in Granular Sands." *Journal of Experimental Biology*, January, jeb.167759. https://doi.org/10.1242/jeb.167759.

Dorgan, Kelly M., Sanjay R. Arwade, and Peter A. Jumars. 2007. "Burrowing in Marine Muds by Crack Propagation: Kinematics and Forces." *Journal of Experimental Biology* 210 (23): 4198–4212. https://doi.org/10.1242/jeb.010371.

Durgunoglu, H Turan, and James K Mitchell. 1973. "Static Penetration Resistance of Soils."

Escobar, Esteban, Miguel Benz Navarrete, Roland Gourvès, Younes Haddani, Pierre Breul, and Bastien Chevalier. 2016. "Dynamic Characterization of the Supporting Layers in Railway Tracks Using the Dynamic Penetrometer Panda 3®." *Procedia Engineering* 143: 1024–33. https://doi.org/10.1016/j.proeng.2016.06.099.

Evangelista, Dennis, Scott Hotton, and Jacques Dumais. 2011. "The Mechanics of Explosive Dispersal and Self-Burial in the Seeds of the Filaree, *Erodium* Cicutarium (Geraniaceae)." *Journal of Experimental Biology* 214 (4): 521–29. https://doi.org/10.1242/jeb.050567.

Gans, Carl. 1968. "Relative Success of Divergent Pathways in Amphisbaenian Specialization." *The American Naturalist* 102 (926): 345–62. https://doi.org/10.1086/282548.

Gayathri, V. L., and Prashanth Vangla. 2024. "Shear Behaviour of Snakeskin-Inspired Ribs and Soil Interfaces." *Acta Geotechnica* 19 (3): 1397–1419. https://doi.org/10.1007/s11440-023-02009-w.

Gray, J., and H. W. Lissmann. 1950. "The Kinetics of Locomotion of the Grass-Snake." *Journal of Experimental Biology* 26 (4): 354–67. https://doi.org/10.1242/jeb.26.4.354.

Huang, Sichuan, and Junliang Tao. 2020. "Modeling Clam-Inspired Burrowing in Dry Sand Using Cavity Expansion Theory and DEM." *Acta Geotechnica* 15 (8): 2305–26. https://doi.org/10.1007/s11440-020-00918-8.

Hunt, O. M., K. B. O'Hara, Y. Chen, and A. Martinez. 2023. "Numerical and Physical Modeling of the Effect of the Cone Apex Angle on the Penetration Resistance in Coarse-Grained Soils." *International Journal of Geomechanics* 23 (2): 04022273. https://doi.org/10.1061/(ASCE)GM.1943-5622.0002626.

Jaeger, Robert Andrew. 2012. *Numerical and Experimental Study on Cone Penetration in Sands and Intermediate Soils*. University of California, Davis.

Jamiolkowski, Michele. 2012. "Role of Geophysical Testing in Geotechnical Site Characterization." *Soils and Rocks* 35 (2): 117–37. https://doi.org/10.28927/SR.352117.

Jung, W., W. Kim, and H.-Y. Kim. 2014. "Self-Burial Mechanics of Hygroscopically Responsive Awns." *Integrative and Comparative Biology* 54 (6): 1034–42. https://doi.org/10.1093/icb/icu026.

Kim, Hye-jeong, Akie Kobayashi, Nobuharu Fujii, Yutaka Miyazawa, and Hideyuki Takahashi. 2016. "Gravitropic Response and Circumnutation in Pea (Pisum Sativum) Seedling Roots." *Physiologia Plantarum* 157 (1): 108–18. https://doi.org/10.1111/ppl.12406.

Kong, X. Q., and C. W. Wu. 2009. "Measurement and Prediction of Insertion Force for the Mosquito Fascicle Penetrating into Human Skin." *Journal of Bionic Engineering* 6 (2): 143–52. https://doi.org/10.1016/S1672-6529(08)60111-0.

Koumoto, T, and GT Houlsby. 2001. "Theory and Practice of the Fall Cone Test." *Géotechnique* 51 (8): 701–12.

Kulić, I.M., M. Mani, H. Mohrbach, R. Thaokar, and L. Mahadevan. 2009. "Botanical Ratchets." *Proceedings of the Royal Society B: Biological Sciences* 276 (1665): 2243–47. https://doi.org/10.1098/rspb.2008.1685.

Lehane, Barry, Zhongqiang Liu, Eduardo Bittar, Farrokh Nadim, Suzanne Lacasse, RJ Jardine, Pasquale Carotenuto, P Jeanjean, M Rattley, and K Gavin. 2020. "A New'unified'CPT-Based Axial Pile Capacity Design Method for Driven Piles in Sand." In , 463–77. American Society of Civil Engineers.

Ling, Jintian, Lelun Jiang, Keyun Chen, Chengfeng Pan, Yan Li, Wei Yuan, and Liang Liang. 2016. "Insertion and Pull Behavior of Worker Honeybee Stinger." *Journal of Bionic Engineering* 13 (2): 303–11. https://doi.org/10.1016/S1672-6529(16)60303-7.

Lobo-Guerrero, Sebastian, and Luis E. Vallejo. 2007. "Influence of Pile Shape and Pile Interaction on the Crushable Behavior of Granular Materials around Driven Piles: DEM Analyses." *Granular Matter* 9 (3–4): 241. https://doi.org/10.1007/s10035-007-0037-3.

Ma, Yifei, T. Matthew Evans, and Douglas D. Cortes. 2020. "2D DEM Analysis of the Interactions between Bio-Inspired Geo-Probe and Soil during Inflation–Deflation Cycles." *Granular Matter* 22 (1): 11. https://doi.org/10.1007/s10035-019-0974-7.

MacDonald, Ian. 2015. *Burial Mechanics of the Pacific Sandfish: The Role of the Ventilatory Pump and Physical Constraints on the Behavior*. Northern Arizona University.

Mak, TW, and LH Shu. 2004. "Abstraction of Biological Analogies for Design." *CIRP Annals* 53 (1): 117–20.

Martinez, A., J.T. DeJong, R.A. Jaeger, and A. Khosravi. 2020. "Evaluation of Self-Penetration Potential of a Bio-Inspired Site Characterization Probe by Cavity Expansion Analysis." *Canadian Geotechnical Journal* 57 (5): 706–16. https://doi.org/10.1139/cgj-2018-0864.

Martinez, Alejandro, Jason Dejong, Idil Akin, Ali Aleali, Chloe Arson, Jared Atkinson, Paola Bandini, et al. 2022. "Bio-Inspired Geotechnical Engineering: Principles, Current Work, Opportunities and Challenges." *Géotechnique* 72 (8): 687–705. https://doi.org/10.1680/jgeot.20.P.170.

Martinez, Alejandro, Sophia Palumbo, and Brian D. Todd. 2019. "Bioinspiration for Anisotropic Load Transfer at Soil–Structure Interfaces." *Journal of Geotechnical and Geoenvironmental Engineering* 145 (10): 04019074. https://doi.org/10.1061/(ASCE)GT.1943-5606.0002138.

Martinez, Alejandro, and Junliang Tao. 2024. "Editorial for Special Issue on Bio-Inspired Geotechnics." *Acta Geotechnica* 19 (3): 1137–38. https://doi.org/10.1007/s11440-024-02323-x.

Martinez, Alejandro, Fabian Zamora, and Daniel Wilson. 2024. "Field Evaluation of the Installation and Pullout of Snakeskin-Inspired Anchorage Elements." *Journal of Geotechnical and Geoenvironmental Engineering*. https://doi.org/10.1061/JGGEFK/GTENG-12311.

Marvi, Hamidreza, and David L. Hu. 2012. "Friction Enhancement in Concertina Locomotion of Snakes." *Journal of The Royal Society Interface* 9 (76): 3067–80. https://doi.org/10.1098/rsif.2012.0132.

Mayne, Paul W. 2014. "Interpretation of Geotechnical Parameters from Seismic Piezocone Tests." In *3rd International Symposium on Cone Penetration Testing (CPT'14)*. Las Vegas, NV: ISSMGE Technical Committee TC 102.

Merz, Rachel Ann, and Deirdre Renee Edwards. 1998. "Jointed Setae – Their Role in Locomotion and Gait Transitions in Polychaete Worms." *Journal of Experimental Marine Biology and Ecology* 228 (2): 273–90. https://doi.org/10.1016/S0022-0981(98)00034-3.

Montana, Jasper, Julian K. Finn, and Mark D. Norman. 2015. "Liquid Sand Burrowing and Mucus Utilisation as Novel Adaptations to a Structurally-Simple Environment in Octopus Kaurna Stranks, 1990." *Behaviour* 152 (14): 1871–81. https://doi.org/10.1163/1568539X-00003313.

Naclerio, Nicholas D., Andras Karsai, Mason Murray-Cooper, Yasemin Ozkan-Aydin, Enes Aydin, Daniel I. Goldman, and Elliot W. Hawkes. 2021. "Controlling Subterranean Forces Enables a Fast, Steerable, Burrowing Soft Robot." *Science Robotics* 6 (55): eabe2922. https://doi.org/10.1126/scirobotics.abe2922.

Naziri, Saeedeh, Cyrena Ridgeway, Jose A. Castelo, Salvador Ibarra, Katarina Provenghi, and Douglas D. Cortes. 2024. "Earthworm-Inspired Subsurface Penetration Probe for Landed Planetary Exploration." *Acta Geotechnica* 19 (3): 1267–74. https://doi.org/10.1007/s11440-024-02240-z.

O'Hara, Kyle B., and Alejandro Martinez. 2022. "Load Transfer Directionality of Snakeskin-Inspired Piles during Installation and Pullout in Sands." *Journal of Geotechnical and Geoenvironmental Engineering* 148 (12): 04022110. https://doi.org/10.1061/(ASCE)GT.1943-5606.0002929.

Ortiz, Daniel, Nick Gravish, and Michael T. Tolley. 2019. "Soft Robot Actuation Strategies for Locomotion in Granular Substrates." *IEEE Robotics and Automation Letters* 4 (3): 2630–36. https://doi.org/10.1109/LRA.2019.2911844.

Passini, Larissa De Brum, Fernando Schnaid, Marcelo Maia Rocha, and Sergio Viçosa Möller. 2018. "Mechanism of Model Pile Installation by Water Jet Fluidization in Sand." *Ocean Engineering* 170 (December): 160–70. https://doi.org/10.1016/j.oceaneng.2018.10.017.

Passini, Larissa De Brum, and Fernando Schnaid. 2015. "Experimental Investigation of Pile Installation by Vertical Jet Fluidization in Sand." *Journal of Offshore Mechanics and Arctic Engineering* 137 (4): 042002. https://doi.org/10.1115/1.4030707.

Pestana, Juan M., and Andrew J. Whittle. 1999. "Formulation of a Unified Constitutive Model for Clays and Sands." *International Journal for Numerical and Analytical Methods in Geomechanics* 23 (12): 1215–43. https://doi.org/10.1002/(SICI)1096-9853(199910)23:12<1215::AID-NAG29>3.0.CO;2-F.

Purdy, Christopher M., Alena J. Raymond, Jason T. DeJong, Alissa Kendall, Christopher Krage, and Jamie Sharp. 2022. "Life-Cycle Sustainability Assessment of Geotechnical Site Investigation." *Canadian Geotechnical Journal* 59 (6): 863–77. https://doi.org/10.1139/cgj-2020-0523.

Rollins, Kyle M., Jashod Roy, Adda Athanasopoulos-Zekkos, Dimitrios Zekkos, Sara Amoroso, and Zhenzhong Cao. 2021. "A New Dynamic Cone Penetration Test–Based Procedure for Liquefaction Triggering Assessment of Gravelly Soils." *Journal of Geotechnical and Geoenvironmental Engineering* 147 (12): 04021141. https://doi.org/10.1061/(ASCE)GT.1943-5606.0002686.

Savioli, A., C. Viggiani, and J. C. Santamarina. 2014. "Root-Soil Mechanical Interaction." In *Geo-Congress 2014 Technical Papers*, 3977–84. Atlanta, Georgia: American Society of Civil Engineers. https://doi.org/10.1061/9780784413272.386.

Sharif, Yaseen Umar, Michael John Brown, Matteo Oryem Ciantia, Benjamin Cerfontaine, Craig Davidson, Jonathan Knappett, Gerrit Johannes Meijer, and Jonathan Ball. 2021. "Using Discrete Element Method (DEM) to Create a Cone Penetration Test (CPT)-Based Method to Estimate the Installation Requirements of Rotary-Installed Piles in Sand." *Canadian Geotechnical Journal* 58 (7): 919–35. https://doi.org/10.1139/cgj-2020-0017.

Simmons, Carl, Dieter Söll, and Fernando Migliaccio. 1995. "Circumnutation and Gravitropism Cause Root Waving in Arabidopsis Thaliana." *Journal of Experimental Botany* 46 (1): 143–50. https://doi.org/10.1093/jxb/46.1.143.

Tang, Yong, and Junliang Tao. 2022. "Multiscale Analysis of Rotational Penetration in Shallow Dry Sand and Implications for Self-Burrowing Robot Design." *Acta Geotechnica*, March. https://doi.org/10.1007/s11440-022-01492-x.

Taylor, Isaiah, Kevin Lehner, Erin McCaskey, Niba Nirmal, Yasemin Ozkan-Aydin, Mason Murray-Cooper, Rashmi Jain, et al. 2021. "Mechanism and Function of Root Circumnutation." *Proceedings of the National Academy of Sciences of the United States of America* 118 (8): e2018940118. https://doi.org/10.1073/pnas.2018940118.

Tovar-Valencia, Ruben D., Ayda Galvis-Castro, Rodrigo Salgado, and Monica Prezzi. 2021. "Effect of Base Geometry on the Resistance of Model Piles in Sand." *Journal of Geotechnical and Geoenvironmental Engineering* 147 (3): 04020180. https://doi.org/10.1061/(ASCE)GT.1943-5606.0002472.

Trueman, ER. 1967. "The Dynamics of Burrowing in Ensis (Bivalvia)." *Proceedings of the Royal Society of London. Series B. Biological Sciences* 166 (1005): 459–76.

Vogel, Steven. 2000. *Cats' Paws and Catapults: Mechanical Worlds of Nature and People*. WW Norton & Company.

Wilson, AJ, AW Robards, and MJ Goss. 1977. "Effects of Mechanical Impedance on Root Growth in Barley, Hordeum Vulgare L. II. Effects on Cell Development in Seminal Roots." *Journal of Experimental Botany* 28 (5): 1216–27.

Winter, A G, V, R L H Deits, D S Dorsch, A H Slocum, and A E Hosoi. 2014. "Razor Clam to RoboClam: Burrowing Drag Reduction Mechanisms and Their Robotic Adaptation." *Bioinspiration & Biomimetics* 9 (3): 036009. https://doi.org/10.1088/1748-3182/9/3/036009.

Winter, Amos G., Robin L. H. Deits, and A. E. Hosoi. 2012. "Localized Fluidization Burrowing Mechanics of *Ensis* Directus." *Journal of Experimental Biology* 215 (12): 2072–80. https://doi.org/10.1242/jeb.058172.

Yang, Xiaotong, Ningning Zhang, Rui Wang, Alejandro Martinez, Yuyan Chen, Raul Fuentes, and Jian-Min Zhang. 2024. "A Numerical Investigation on the Effect of Rotation on the Cone Penetration Test." *Canadian Geotechnical Journal*, no. ja.

Zhang, Ningning, Yuyan Chen, Alejandro Martinez, and Raul Fuentes. 2023. "A Bioinspired Self-Burrowing Probe in Shallow Granular Materials." *Journal of Geotechnical and Geoenvironmental Engineering* 149 (9): 04023073. https://doi.org/10.1061/JGGEFK.GTENG-11507.

Zhong, Yi, Sichuan Huang, and Junliang "Julian" Tao. 2023. "Minimalistic Horizontal Burrowing Robots." *Journal of Geotechnical and Geoenvironmental Engineering* 149 (4): 02823001. https://doi.org/10.1061/JGGEFK.GTENG-11468.